\documentclass[aps,reprint,amssymb, amsmath, superscriptaddress, showpacs, footinbib, prb]{revtex4-1}
\usepackage{graphicx,epstopdf}
\usepackage{amsmath}
\begin{document}

\title{A case study of bilayered spin-$1/2$ square lattice compound [VO(HCOO)$_2\cdot$(H$_2$O)]}

\author{S. Guchhait}
\author{U. Arjun}
\affiliation{School of Physics, Indian Institute of Science Education and Research Thiruvananthapuram-695551, India}
\author{P. K. Anjana}
\affiliation{School of Chemistry, Indian Institute of Science Education and Research Thiruvananthapuram-695551, India}
\author{M. Sahoo}
\affiliation{Department of Physics, University of Kerala, Kariavattom, Thiruvananthapuram-695581, India}
\author{A. Thirumurugan}
\affiliation{School of Chemistry, Indian Institute of Science Education and Research Thiruvananthapuram-695551, India}
\author{A. Madhi}
\affiliation{School of Physics, Indian Institute of Science Education and Research Thiruvananthapuram-695551, India}
\author{Y. Skourski}
\affiliation{Dresden High Magnetic Field Laboratory (HLD-EMFL), Helmholtz-Zentrum Dresden-Rossendorf, 01328 Dresden, Germany}
\author{B. Koo}
\author{J. Sichelschmidt}
\author{B. Schmidt}
\author{M. Baenitz}
\affiliation{Max Planck Institute for Chemical Physics of Solids, Nthnitzer Str. 40, 01187 Dresden, Germany}
\author{R. Nath}
\email{rnath@iisertvm.ac.in}
\affiliation{School of Physics, Indian Institute of Science Education and Research Thiruvananthapuram-695551, India}
\date{\today}

\begin{abstract}
We present the synthesis and a detail investigation of structural and magnetic properties of polycrystalline [VO(HCOO)$_2\cdot$(H$_2$O)] by means of x-ray diffraction, magnetic susceptibility, high-field magnetization, heat capacity, and electron spin resonance measurements. It crystallizes in a orthorhombic structure with space group $Pcca$. It features distorted VO$_6$ octahedra connected via HCOO linker (formate anions) forming a two-dimensional square lattice network with a bilayered structure. Analysis of magnetic susceptibility, high field magnetization, and heat capacity data in terms of the frustrated square lattice model unambiguously establish quasi-two-dimensional nature of the compound with nearest neighbour interaction $J_1/k_{\rm B} \simeq 11.7$~K and next-nearest-neighbour interaction $J_2/k_{\rm B} \simeq 0.02$~K. It undergoes a N\'eel antiferromagnetic ordering at $T_{\rm N} \simeq 1.1$~K. The ratio $\theta_{\rm CW}/T_{\rm N} \simeq 10.9$ reflects excellent two-dimensionality of the spin-lattice in the compound. A strong in-plane anisotropy is inferred from the linear increase of $T_{\rm N}$ with magnetic field, consistent with the structural data.
\end{abstract}

\pacs{75.30.Et, 75.50.Ee, 75.40.Cx, 75.50.-y, 75.10.Jm}
\maketitle

\section{\textbf{Introduction}}
\begin{figure*}
	\includegraphics[width = \linewidth]{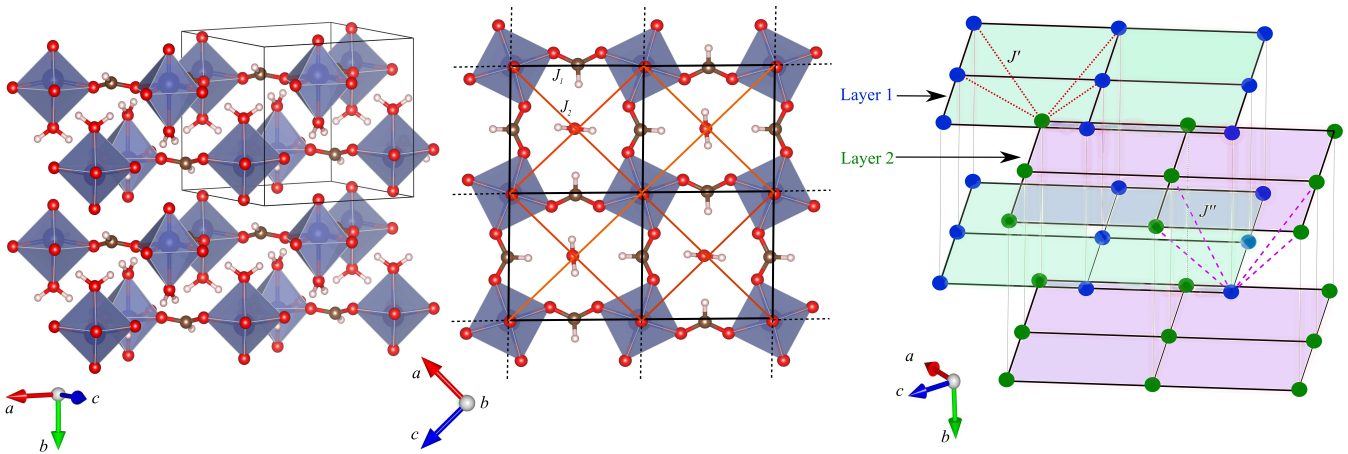}
	\caption{Left panel: Three dimensional view of the crystal structure showing VO(HCOO)$_2$ double layers lying perpendicular to the $b$-axis. Middle panel: A section of one layer in the $ac$-plane, showing square lattice of VO$_6$ octahedra connected through HCOO bridges and the frustrated square lattice or the $J_1 - J_2$ model. Right panel: Spin lattice showing two layers and possible exchange couplings between the layers.}
	\label{Fig_1}
\end{figure*}
Over the past decades, two-dimensional (2D) antiferromagnetic (AFM) spin-$1/2$ systems have played an important role to understand the phase transitions and critical phenomena in magnetic materials.\cite{Manousakis1} The thermodynamic properties of such systems are now a days well established by extensive numerical studies.\cite{Makivic3562,Kim2705,Sandvik11678} The ideal 2D Heisenberg antiferromagnets conventionally lack long range order (LRO) down to zero temperature, following the Mermin-Wagner theorem.\cite{Mermin1133} But real materials inevitably posses a non-negligible inter-plane coupling that triggers the LRO at a finite temperature.\cite{Yasuda217201} When the inter-plane couplings are frustrated and inactive, the LRO is driven by anisotropy terms in the spin Hamiltonian. In addition to the frustrated inter-plane couplings, competing interactions [e.g. nearest-neighbour interaction ($J_1$) along the edge with next-nearest-neighbour interaction ($J_2$) along the diagonal of the square] in spin-$1/2$ 2D systems, known as $J_1 - J_2$ model, often destabilize LRO, giving rise to various novel non-magnetic ground states.\cite{Shannon599,Shannon027213,Nath064422,Nath214430,Tsyrulin197201}
Even in high-$T_{\rm c}$ cuprates, this 2D AFM correlations are believed to be an essential ingredient for superconductivity.\cite{Manousakis1,Lee012501} Most interestingly, a recent report by Jain $et~al$.\cite{Jain2017} suggests that a condensed-matter analog for the physics of Higgs boson decay, which is very important in particle physics, can be provided by 2D AFM materials. It is anticipated that condensed matter realization of Higgs boson can provide insights regarding its behaviour in different symmetries and dimensionalities.\cite{Pekker269}

Though, there are numerous experimental studies on spin-$1/2$ square lattices, majority of them are focused on purely inorganic systems and only few studies are reported on metal-organic based materials. The advantage of these metal-organic systems is that one can tune their physical properties simply by changing the organic ligands.\cite{Goddard077208} Secondly, metal-organic complexes have relatively weak exchange couplings compared to the inorganic compounds which makes them promising candidates for high field experiments, especially to explore the field induced quantum phenomenon.\cite{Nath054409,Rnnow037202,Tsyrulin197201} An interesting example of this category is Cu(pz)$_2$(ClO$_4$)$_2$ which has attracted a lot more attention experimentally as well theoretically.\cite{Tsyrulin134409,Tsyrulin197201,Siahatgar064431,Lancaster094421}
Layered metal-organic complexes also exhibit various peculiar electronic properties such as metal-insulator transition, Fermi liquid behaviour, unconventional superconductivity etc.\cite{Lefebvre5420,Nam584,Ishiguro2006,Manna016403} Unlike inorganic compounds, the mechanism behind all these phenomenon in organic based metal complexes are not yet understood. Therefore, in recent days, there is an enduring demand to synthesize metal-organic based spin-$1/2$ 2D model compounds and investigate the physical properties in order to elucidate their relevance in strongly correlated physics.

In this paper, we present the magnetic properties of a new spin-$1/2$ quasi-2D AFM compound, [VO(HCOO)$_2\cdot$(H$_2$O)] investigated via magnetic susceptibility, high field magnetization, heat capacity, and electron spin resonance (ESR) measurements. The vanadyl formate [VO(HCOO)$_2\cdot$(H$_2$O)] crystallizes in a orthorhombic structure with space group $Pcca$.\cite{Mootz1218} As shown in the left panel of Fig.~\ref{Fig_1}, the unit cell contains two VO(HCOO)$_2$ layers which are interdigitated by co-ordinated water molecules to form a bilayer system. The bilayers are repeated along the $b$-direction to form a three-dimensional(3D) structure. In each VO(HCOO)$_2$ layer, a 2D square lattice network is formed by the distorted VO$_6$ octahedra linked via HCOO bridges. The distance between V$^{4+}$ ions along the edges of the square is found to be $\sim 5.977$~\AA and are coupled with an exchange coupling $J_1$ as shown in the middle panel of Fig.~\ref{Fig_1}. There is no visible connecting path among the V$^{4+}$ ions along the diagonals of the square. Therefore, the interaction along the diagonals ($J_2$) is expected to be very weak. It was further observed that the distance between the V$^{4+}$ ions along the diagonals of the square are different i.e. $\sim 8.395$~\AA and $\sim 8.51$~\AA. This inequality may produce a strong in-plane anisotropy in the compound. Similarly, there is no bonding between the inter-layer vanadyl groups and hence, the inter-layer interactions if at all present, should be negligible compared to the intra-layer interactions.

The right panel of Fig.~\ref{Fig_1} presents a schematic view of the spin-lattice with possible exchange network between the layers. One can see that vanadyl groups of layer 1 may interact with that of layer 2 (separated by a distance of $ \sim 5.41 $~\AA) in a triangular fashion with exchange coupling $J^{\prime}$. Similarly vanadyl group of layer 2 may interact with that of next layer 1 (separated by a distance of $ \sim 5.86 $~\AA) in a triangular fashion with exchange coupling $J^{\prime\prime}$. The frustrated triangular network between the layers is expected to suppress the LRO to very low temperatures, retaining the two-dimensionality of the system over a large temperature range. Thus, the bilayered nature and the frustrated triangular inter-layer interactions make [VO(HCOO)$_2\cdot$(H$_2$O)] an unusual candidate compared to other 2D compounds.

Our magnetic measurements reveal that it is a spin-$1/2$ square lattice compound with $J_1 \simeq 11.7$~K and $J_2 \simeq 0.02$~K. It undergoes a N\'eel AFM ordering at $T_{\rm N} \simeq 1.1$~K. The ratio $\theta_{\rm CW}/T_{\rm N}$ is found to be quite large, making it convenient to investigate its magnetic properties over a wide range of temperature.

\section{\textbf{Methods}}
\begin{figure}
	\includegraphics{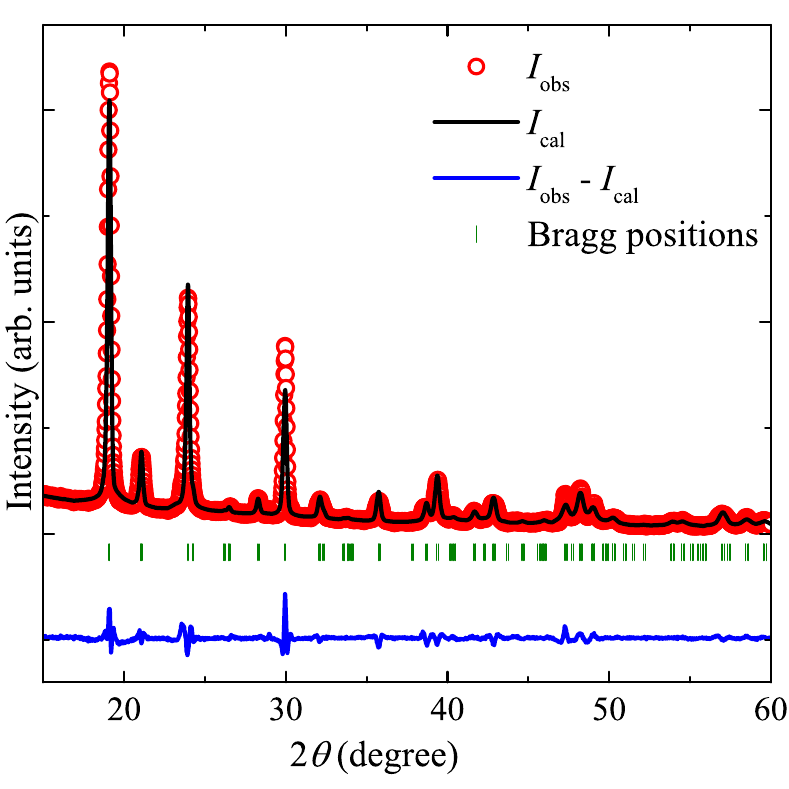}
	\caption{(Color online) Powder XRD pattern (open circles) at room temperature for [VO(HCOO)$_2\cdot$(H$_2$O)]. The solid line is the Le-Bail fit, the vertical bars represent the expected Bragg peak positions, and the lower solid line corresponds to the difference between the observed and calculated intensities.}
	\label{Fig_2}
\end{figure}

Synthesis of [VO(HCOO)$_2\cdot$(H$_2$O)] was performed following the conventional solvothermal route. In a typical reaction, 0.163~g (1~mmol) of VOSO$_4\cdot x $H$_2$O (Aldrich, 97$\%$) and 6~ml (159~mmol) of formic acid (Spectrochem, 98$\%$) were mixed and heated at 100~$^\circ$C for 3~days in a teflon-lined stainless steel bomb of internal volume 20~mL. The resulting product was found to be light blue powder that consists of plate shaped crystals of the title compound. Single crystal x-ray diffraction (XRD) on a good quality single crystal confirms the orthorhombic ($Pcca$) crystal structure of the compound.\cite{Gilson136} To further cross check the phase purity, powder XRD was performed on the crushed powder sample at room temperature using a PANalytical (Cu $K_{\alpha}$ radiation, $\lambda_{\rm ave} = 1.54182$~\AA) powder diffractometer. Le-Bail fit of the powder XRD pattern was performed using FullProf package\cite{Carvajal55} taking the initial structural parameters from Ref.~[\onlinecite{Gilson136}]. Figure~\ref{Fig_2} presents the powder XRD pattern of [VO(HCOO)$_2\cdot$(H$_2$O)] at room temperature along with the fit. All the peaks could be fitted using the orthorhombic ($Pcca$) structure. The obtained best fit parameters are $a = 8.434(1)$~\AA, $b = 7.4336(8)$~\AA, $c = 8.4418(9)$~\AA, and the goodness-of-fit $\chi^2 \simeq 6.68$. These lattice parameters are consistent with the earlier report.~\cite{Gilson136}

Magnetic susceptibility $\chi$ was measured as a function of temperature (0.5~K~$\leq T \leq$~380~K) and applied magnetic field $H$. In the high temperature range ($T \geq 2$~K), measurements were done using the vibrating sample magnetometer (VSM) attachment to the Physical Property Measurement System [PPMS, Quantum Design]. For $T \leq 2$~K, measurements were carried out using a $^3$He attachment to the SQUID magnetometer (MPMS-7, Quantum Design). High field magnetization ($M$ vs $H$) was measured at $T = 1.5$~K in pulsed magnetic field upto 40~T at the Dresden High magnetic field Laboratory (HLD). Heat capacity, $C_{\rm p}(T)$ was also measured using heat capacity option of the PPMS on a sintered pellet. For the low temperature (0.35~K~$\leq T \leq 2$~K) $C_{\rm p}$ measurements, an additional $^{3}$He attachment was used in the PPMS. Since the size of the single crystals were very small, all the measurements were carried out on the powder sample, obtained by crushing a large number of single crystals.

The ESR experiments were carried out on a powder sample with a standard continuous-wave spectrometer in the temperature range 3~K~$\leq T \leq 300$~K. The power $P$ absorbed by the sample from a transverse magnetic microwave field ($X$-band, $\nu\simeq 9.4$\,GHz) was measured as a function of the external magnetic field $H$. The final data were recorded as the derivative $dP/dH$ with $H$. The ESR $g$-factor was estimated using the resonance condition $g=\frac{h\nu}{\mu_{\rm B} H_{\rm res}}$, where $h$ is the Planck's constant, $\mu_{\rm B}$ is the Bohr magneton, $ \nu $ is the resonance frequency, and $ H_{\rm res} $ is the corresponding resonance field.

For comparison with theory, quantum Monte Carlo (QMC) simulation for magnetization was performed assuming the Heisenberg model on a non-frustrated square lattice with Hamiltonian ${\cal H} = J\sum_{ij}\vec{S}_i\cdot\vec{S}_j - H\sum_iS^z_i$, where $J$ is the exchange coupling between spins at the $i$th and $j$th sites and $H$ is magnetic field strength. We used the ALPS\cite{ALPS} code for the directed loop QMC algorithm in the stochastic series expansion representation.\cite{SandvikR14157,*Alet036706,*Pollet056705} The lattice size was taken to be $40\times 40$. We typically did $10^5$ sweeps including around $5000$ number of thermalization sweeps.

\section{\textbf{Results}}
\subsection{\textbf{Magnetization}}
\begin{figure}
	\includegraphics [width = \linewidth]{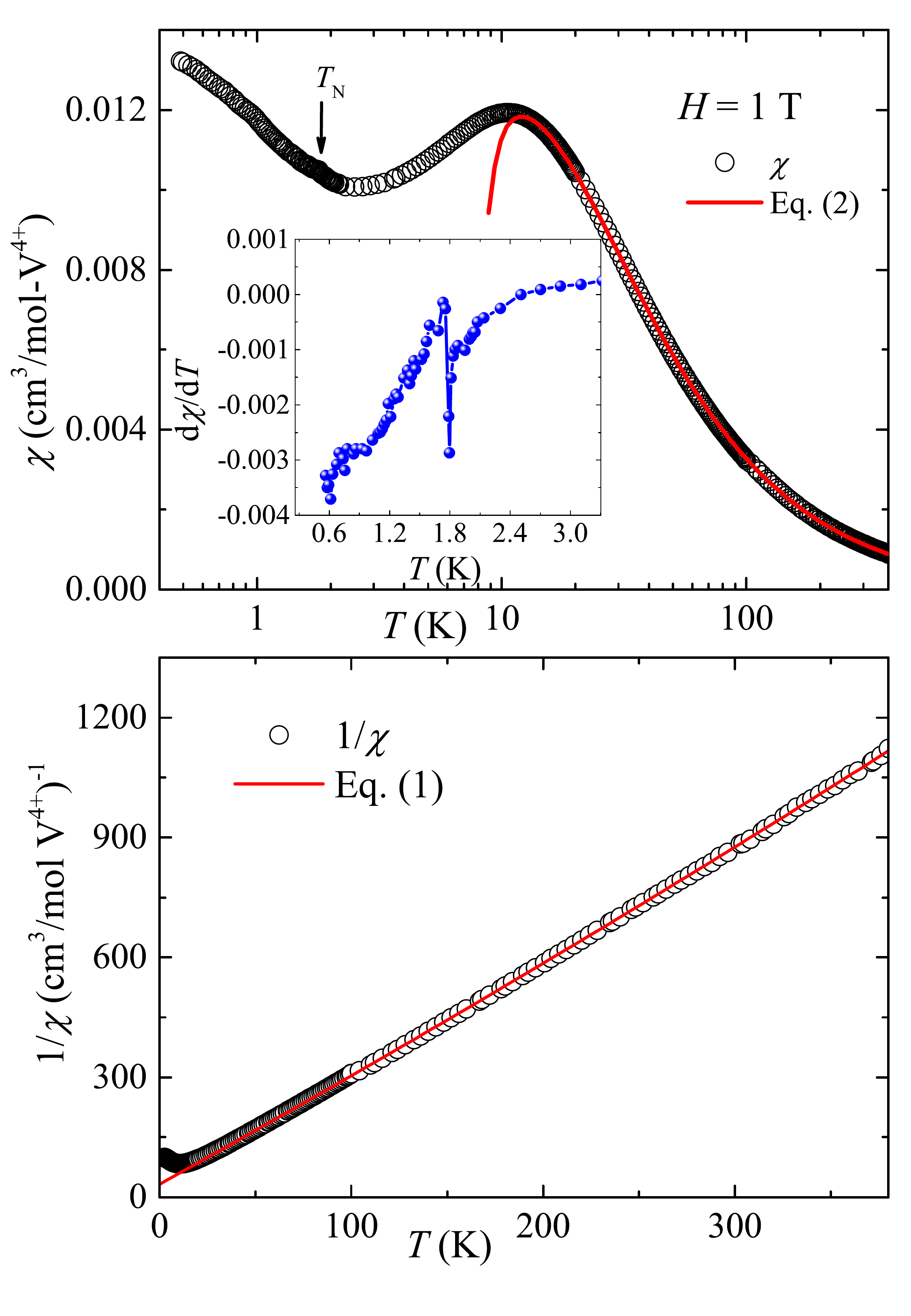}
	\caption{ Upper panel: $\chi(T)$ measured at an applied field of $H=1$~T. Solid line represents the fit using 2D frustrated square lattice model [Eq.~(\ref{chi_twoD})]. Inset: $d\chi/dT$ vs $T$ in the low temperature regime. Lower panel: Inverse magnetic susceptibility ($1/\chi$) at $H=1$~T as a function of $T$. Solid line is the fit by Eq.~(\ref{cw}).}
	\label{Fig_3}
\end{figure}

Temperature dependent magnetic susceptibility $\chi(T)$ measured at an applied field of $H = 1$~T is shown in the upper panel of Fig.~\ref{Fig_3}. As the temperature is lowered, $\chi(T)$ increases in a Curie-Weiss manner and then shows a broad maximum ($T_\chi^ {\rm {max}}$) at about 10~K. This broad maximum is suggestive of a short range magnetic order which is also a hallmark of low dimensionality. It exhibits a weak cusp at $T_{\rm N} \simeq 1.5 $~K, a possible indication of the occurrence of a magnetic LRO. With further reduction in temperature, a small upturn was observed which is likely due to the defects present in the sample. The inset of Fig.~\ref{Fig_3} shows $d\chi/dT$ vs $T$ in the low temperature regime, which clearly indicates the presence of a magnetic LRO at the $ T_{\rm N}$.

For extracting the magnetic parameters, $\chi(T)$ at high temperatures was fitted by the following expression
\begin{equation}\label{cw}
\chi(T) = \chi_0 + \frac{C}{T + \theta_{CW}},
\end{equation}
where $\chi_0$ is the temperature independent susceptibility consisting of core diamagnetism of the core electron shells and Van-Vleck paramagnetism of the open shells of the V$^{4+}$ ions in the sample.
The second term in Eq.~\eqref{cw} is the Curie-Weiss (CW) law with the Curie-Weiss temperature ($\theta_{\rm CW}$) and Curie constant $C = N_{\rm A} \mu_{\rm eff}^2/3k_{\rm B}$, where $N_{\rm A}$ is Avogadro's number, $k_{\rm B}$ is Boltzmann constant, $\mu_{\rm eff} = g\sqrt{S(S+1)} \mu_{\rm B}$ is the effective magnetic moment, $g$ is the Land$\acute{\rm e}$ $g$-factor, $\mu_{\rm B}$ is the Bohr magneton, and $S$ is the spin quantum number. Our fit in the temperature range 110~K to 380~K (lower panel of Fig.~\ref{Fig_3}) yields $\chi_0 \simeq -6.363 \times 10^{-5}$~cm$^3$/mol-V$^{4+}$, $C \simeq 0.376$~cm$^3$K/mol-V$^{4+}$, and $\theta_{\rm CW} \simeq$~12~K. From the value of $C$, the effective moment is calculated to be $\mu_{\rm eff} \simeq 1.73~ \mu_{\rm B}$/V$^{4+}$ which exactly matches with the expected spin-only value for $S = 1/2$ with $g = 2$. The positive value of $\theta_{\rm CW}$ is indicative of the antiferromagnetic exchange interaction among the V$^{4+}$ ions.\cite{Domb1964}

\begin{figure}
	\includegraphics[width = \linewidth]{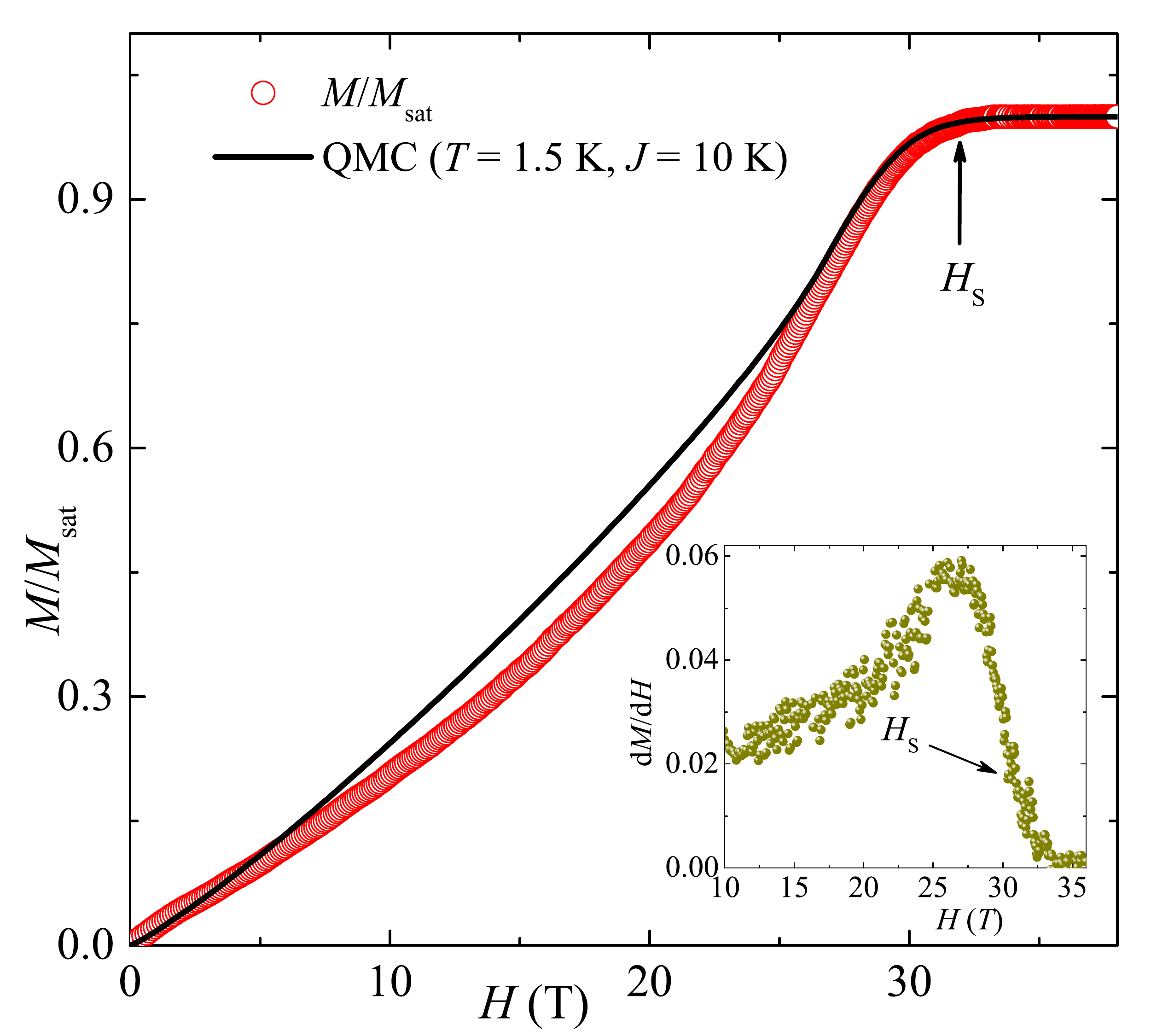}\\
	\caption{Magnetization (normalized to one) vs field measured at $T = 1.5$~K. The solid line represents the QMC simulation, assuming a uniform square lattice model. Inset: $dM/dH$ vs $H$ highlighting the saturation field.}
	\label{Fig_4}
\end{figure}
To understand the exchange network, the experimental $\chi(T)$ data were fitted by the equation
\begin{equation}
\label{chi_twoD}
\chi(T) = \chi_0 + \chi_{\rm spin}(T).
\end{equation}
Here, $\chi_{\rm spin}(T)$ is the high temperature series expansion (HTSE) of spin susceptibility for the spin-$1/2$ frustrated square lattice (FSL) or $J_1 - J_2$ model.\cite{Rosner014416,Schmidt104443} The expression is given by
\begin{eqnarray}
	\label{chi_spin}
	\chi_{\rm spin}(T)=\frac{N_{A}g^{2}\mu_{B}^{2}}{k_{B}T}\sum_{n}\left(\frac{J_{1}}{k_{B}T}\right) ^{n}\sum_{m}c_{m,n}\left(\frac{J_{2}}{J_{1}}\right)^{m},
\end{eqnarray}
where, $c_{m,n}$ are the coefficients listed in Table~I of Ref.~[\onlinecite{Rosner014416}].
This HTSE is valid only in the high temperature region $T\geq J_i$.
We fitted the experimental $\chi(T)$ data by Eq.~\eqref{chi_twoD} in the temperature range 12~K to 380~K fixing $g = 2$, obtained from the ESR experiments. It yields two solutions with equally good fits. Solution I: $\chi_0 \simeq -6.846 \times 10^{-5}$~cm$^3$/mol-V$^{4+}$, $J_1/k_{\rm B} \simeq 11.7$~K, and $J_2/k_{\rm B} \simeq 0.02$~K and Solution II: $\chi_0 \simeq -6.936 \times 10^{-5}$~cm$^3$/mol-V$^{4+}$, $J_1/k_{\rm B} \simeq 11.7$~K, and ferromagnetic $J_2/k_{\rm B} \simeq -0.02$~K. In both the solutions, the magnitude of $J_2$ is almost three orders of magnitude smaller or almost negligible compared to $J_1$, as is expected from the structural data. It also implies that the system can be viewed as a non-frustrated square lattice. Since for a FSL, $\theta_{\rm CW} = J_1 + J_2$, our experimentally obtained higher value of $\theta_{\rm CW}$ favours solution-I with AFM $J_1$ and $J_2$. Nevertheless, in both the cases the $J_1$ and $J_2$ values locate the system in the N\'eel antiferromagnetic (NAF) region of the $J_1 - J_2$ phase diagram.\cite{Shannon599}

As it is observed from the zero-field $C_{\rm p}(T)$ data, the compound undergoes a magnetic LRO at $ T_{\rm N} \simeq 1.1 $~K. This suggests that there are non-negligible inter-plane couplings in contrast to what was expected from the structural data. From the value of $T_{\rm N}$, $J_1$, and $J_2$, one can calculate the average inter-plane coupling $J_\perp$ using the relation\cite{Majlis7872,Schmidt214443}
\begin{equation}
\label{interJ}
k_{\rm B}T_{\rm N} \simeq \pi(J_1 - 2J_2)/[2 + ln((J_1-2J_2)/J_\perp)],
\end{equation}
where non-frustrated inter-layer couplings are assumed. Taking the appropriate values ($T_{\rm N} \simeq 1.1$~K, $J_1/k_{\rm B} \simeq 11.7$~K, and $J_2/k_{\rm B} \simeq 0.02$~K), $J_\perp/k_{\rm B}$ is calculated to be $J_\perp/k_{\rm B} \simeq 3.3 \times 10^{-13}$~K. This value of $J_\perp/k_{\rm B}$ is found to be unrealistically low and even several orders of magnitude smaller than the dipole-dipole coupling. Such a discrepancy could be due to the bilayer nature of the spin-lattice and the presence of inter-layer frustration.

In order to check whether there is any field induced effects and to obtain the saturation magnetization, high field magnetization was measured at $T = 1.5$~K upto 40~T. Figure~\ref{Fig_4} presents the magnetization ($M$) vs $H$ normalized to 1. At the low field regime, $M$ increases almost linearly with $H$ and then shows a pronounced curvature before it saturates completely at $H_{\rm S} \simeq 32$~T. Such a pronounced curvature is indicative of strong quantum fluctuations or frustration in the spin system. The inset of Fig.~\ref{Fig_4} presents the derivative $dM/dH$ as a function of $H$ to magnify the change in slope at the saturation field $H_{\rm S}$.
To reconfirm the magnitude of exchange couplings we analyzed the value of the saturation field $H_{\rm S}$. According to theoretical results by Schmidt~\textit{et al.},\cite{Schmidt125113} the saturation field in a FSL model can be calculated as
\begin{equation}
\begin{split}
H_{S}=\frac{J_{c}k_{B}zS}{g\mu _{B}} & \left[\left( 1-\frac{1}{2}(\cos
Q_{x}+\cos Q_{y})\right) \cos \varphi\right. \\
&\quad \left. {}+(1-\cos Q_{x}\cos Q_{y})\sin \varphi\vphantom{\frac12}\right],
\end{split}
\end{equation}
where $z=4$ (magnetic coordination number), $S=1/2$, angle $\phi =$$\text{tan}$$^{-1}(J_{2}/J_{1})$, $J_{\rm c} = \sqrt{J_1^2+J_2^2}$, and $(Q_{x},Q_{y})$
is the wave vector of the ordered state. Using the appropriate wave vectors for the NAF ($\pi$, $\pi$) phase, one can have $H_{S}= 4J_{1}k_B/(g\mu _{B})$. Using this formula, our experimental value of $H_{\rm S} \simeq 32$~T 
corresponds to $J_1/k_{\rm B} \simeq 10.7$~K which is slightly smaller than the one ($\sim 11.7$~K) obtained from the $\chi(T)$ analysis.

\subsection{\textbf{ESR}}
\begin{figure}
	\includegraphics [width = \linewidth]{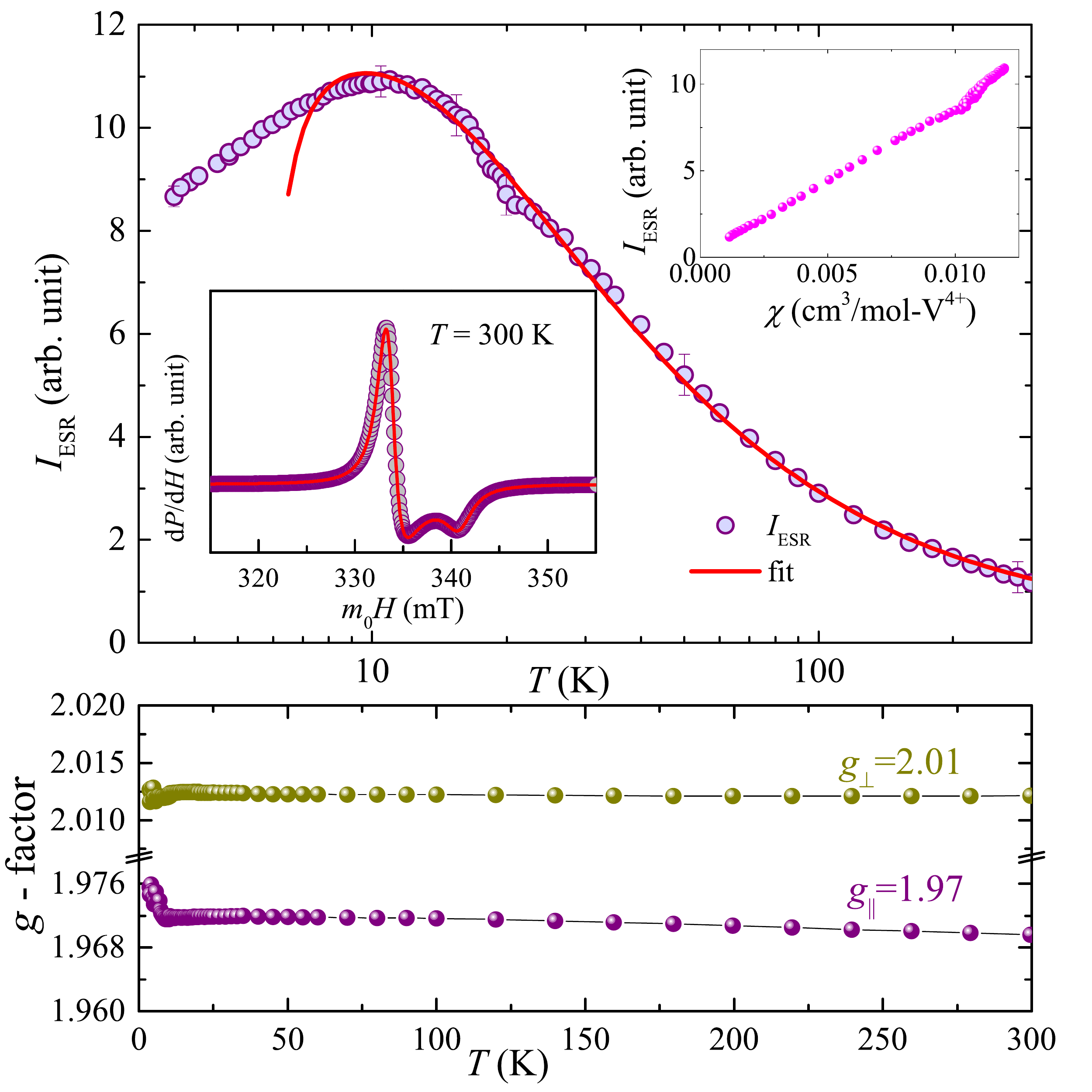}\\
	\caption{Upper panel: Temperature dependent ESR intensity, $I_{\rm ESR}(T)$, obtained by integration of the ESR spectra of the polycrystalline sample. The solid line represents the fit described in the text. Lower inset: a typical spectrum (symbols) together with the fit using a powder averaged Lorentzian shape for a uniaxial $g$-factor anisotropy. Top inset: $I_{\rm ESR}$ vs $\chi$. Lower panel: Temperature dependent $ g $-factor ($g_{\parallel}$ and $g_{\perp}$) obtained from the Lorentzian fit.}
	\label{Fig_5}
\end{figure}
The ESR experimental results on the [VO(HCOO)$_2\cdot$(H$_2$O)] powder sample are presented in Fig.~\ref{Fig_5}. The lower inset of Fig.~\ref{Fig_5} illustrates a typical ESR spectrum at room temperature. We fitted the spectra using a powder-averaged Lorentzian line shape. The fit reproduces the spectral shape very well at $T=300$~K, yielding anisotropic $g$-factors: parallel component $g_{\parallel} \simeq 1.97$ and perpendicular component $g_{\perp} \simeq 2.01$. The isotropic $g$-value
$\left[ =\sqrt{(g^{2}_{\parallel}+2g^{2}_{\perp})/3}\right]$ is calculated to be $g \simeq 2.0$. As shown in the lower panel of Fig.~\ref{Fig_5}, both $g_{\parallel}$ and $g_{\perp}$ are temperature independent in the high temperature range. The line width at half maximum ($ \Delta H $) is also found to be temperature independent at high temperatures. For $T < 8$~K, both $ \Delta H (T)$ and $g(T)$ show a gradual increase reflecting the appearance of spin correlations coming from the magnetic LRO at low temperatures. The ESR intensity ($ I_{\rm ESR} $) as a function of temperature shows a broad maximum at $ \sim 10 $~K similar to the bulk $ \chi(T) $ data. In the upper inset of Fig.~\ref{Fig_5}, $ I_{\rm ESR} $ is plotted as a function of $\chi$. It shows linearity over the whole measured temperature range providing clear evidence that $ I_{\rm ESR}(T)$ probes $ \chi(T) $.

To estimate the exchange coupling, $ I_{\rm ESR}(T)$ data were fitted by the equation
\begin{equation}\label{ESR_twoD}
I_{\rm ESR} = A + B\chi_{\rm spin}(T),
\end{equation}
where $A$ and $B$ are constants and $ \chi_{\rm spin} $ is given in Eq.~\eqref{chi_spin}. The fit in the range 13~K to 380~K (upper panel of Fig.~\ref{Fig_5}) yields $J_1/k_{\rm B} \simeq 10.2$~K and $J_2/k_{\rm B} \simeq 0.007$~K, fixing $g=2$. These values of exchange couplings are close to the ones obtained from the high field data but slightly smaller in magnitude than the ones obtained from the $\chi (T)$ analysis.

\subsection{\textbf{Heat Capacity}}
\begin{figure}
	\includegraphics{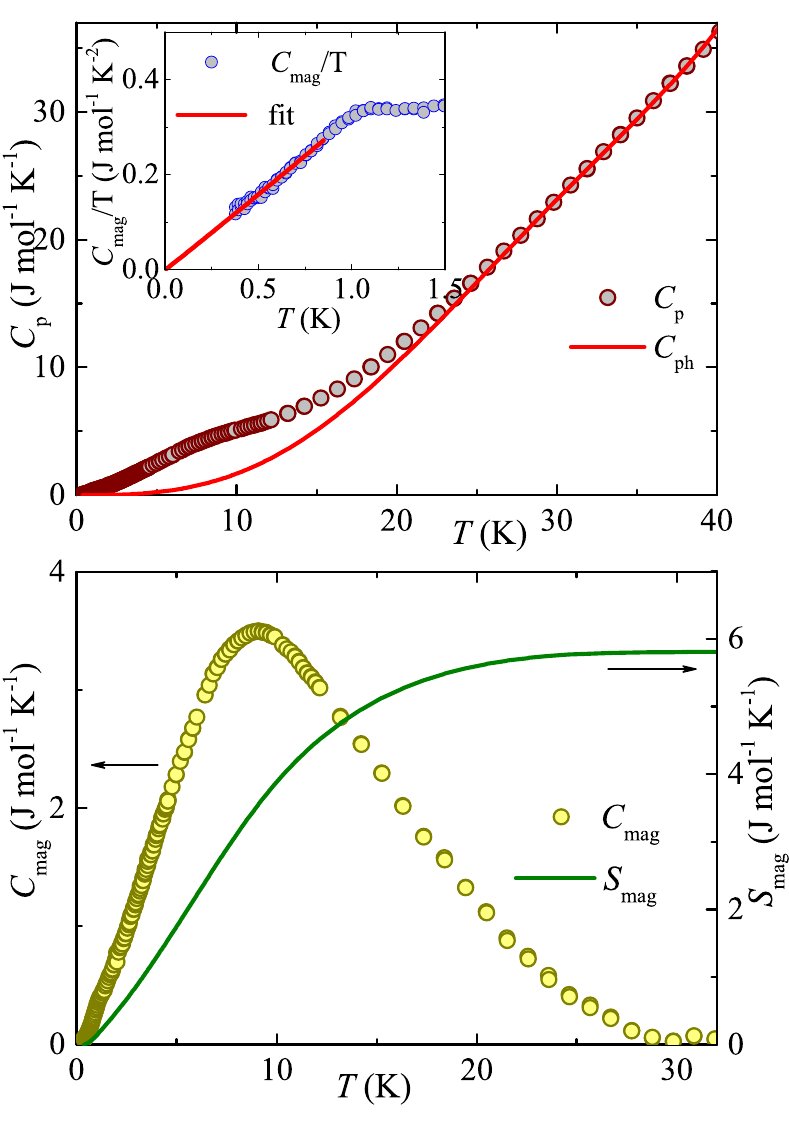}
	\caption{Upper panel: Heat capacity $C_{\rm p}(T)$ of [VO(HCOO)$_2\cdot$(H$_2$O)] measured in zero applied field along with the calculated $C_{\rm ph}(T)$. Inset: $C_{\rm mag}/T$ vs $T$ and the solid line is a linear fit. Lower panel: $C_{\rm mag}$ (left $y$-axis) and $S_{\rm mag}$ (right $y$-axis) are plotted as a function of $T$.}
	\label{Fig_6}
\end{figure}
The heat capacity $C_{\rm p}$ measured in zero applied field is shown in the upper panel of Fig.~\ref{Fig_6}. In a magnetic insulator, $C_{\rm p}$ has two major contributions: one from phonon excitations ($C_{\rm ph}$) and the other one is from the magnetic lattice ($C_{\rm mag}$). At high temperatures, $C_{\rm p}(T)$ is completely dominated by the contribution of $C_{\rm ph}$ while at low temperatures, it is dominated by $C_{\rm mag}$. Our $C_{\rm p}(T)$ data show a weak broad maximum at $T_{\rm C}^{\rm max} \simeq 8.5$~K, similar to that observed in $\chi(T)$. A weak anomaly is detected at around $T_{\rm N} \simeq 1.1$~K associated with the magnetic LRO. With further decrease in $T$, $C_{\rm p}(T)$ decreases gradually towards zero.

In order to estimate the phonon part of the heat capacity, $C_{\rm p}(T)$ data at high temperature $(T > 25$~K) were fitted by the polynomial
\begin{equation}\label{hceq}
C_{\rm ph}(T) = aT^3 + bT^5 + cT^7 + dT^9,
\end{equation}
where $a$, $b$, $c$, and $d$ are arbitrary constants.\cite{polynomial_fit} Similar procedure has been adopted earlier and proven to be an efficient method for the estimation of $C_{\rm ph}$ in the case of metal-organic complexes.\cite{Nath054409,Matsumoto9993,Lancaster094421} The fit was extrapolated down to low temperatures and the $C_{\rm mag}$ was obtained by subtracting the fitted data from the experimental $C_{\rm p}$ data. To crosscheck the reliability of the fitting procedure, we calculated the total change in magnetic entropy $(S_{\rm mag})$ by integrating $C_{\rm mag}(T)/T$ from 0.35~K to high-temperatures as
$S_{\rm mag}(T) = \int_{0.35~K}^{T}\frac{C_{\rm mag}(T')}{T'}dT'$.
The resulting magnetic entropy is $S_{\rm mag} \simeq 5.8$~J/mol~K at 30~K. This value reasonably matches with the expected theoretical value [$S_{\rm mag} = R ln(2S+1)$] of 5.76~J/mol~K for [VO(HCOO)$_2\cdot$(H$_2$O)]. The obtained $C_{\rm mag}(T)$ is presented in the lower panel of Fig.~\ref{Fig_6}. Below $T_{\rm N}$, $C_{\rm mag}(T)$ follows a power law $T^{\alpha}$ with a reduced exponent $\alpha \simeq 2$ (see the inset of the upper panel of Fig.~\ref{Fig_6}).

\begin{figure}
	\includegraphics[scale=1]{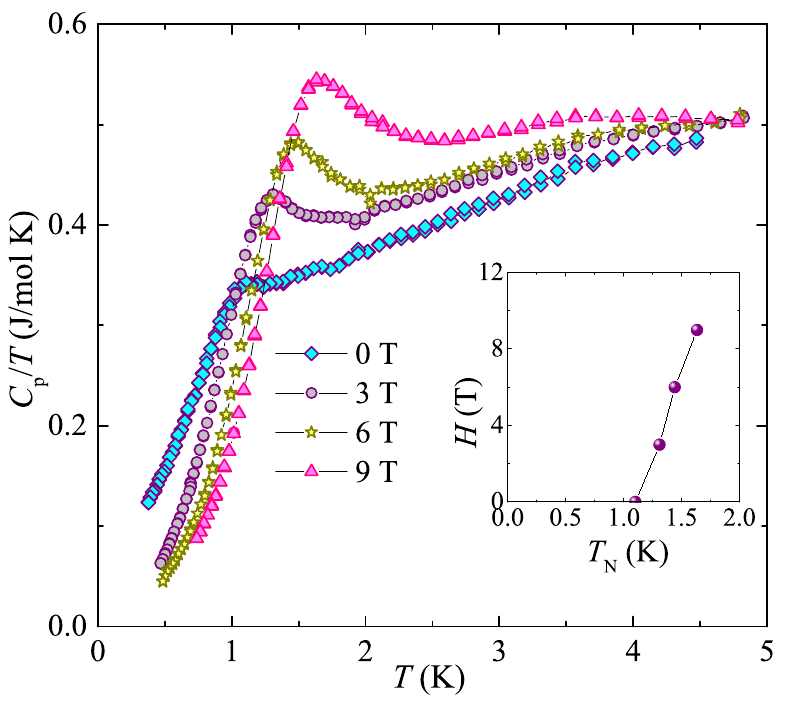}
	\caption{$C_{\rm p}/T$ measured at different applied magnetic fields in the low temperature region. Inset: The variation of $T_{\rm N}$ with $H$.}
	\label{Fig_7}
\end{figure}
To understand the nature of ordering at $T_{\rm N}$, $C_{\rm p}(T)$ of [VO(HCOO)$_2\cdot$(H$_2$O)] was measured at different applied magnetic fields (see Fig.~\ref{Fig_7}). With increasing magnetic field, the peak height at $T_{\rm N}$ is found to be increasing and the peak position is shifting towards higher temperatures. This behaviour is just opposite to what is expected for an AFM 3D ordering. In the inset of Fig.~\ref{Fig_7}, $T_{\rm N}$ vs $H$ is plotted. $T_{\rm N}$ varies almost linearly with $H$ upto the maximum measured field of 9~T.

\section{\textbf{Discussion}}
\begin{table*}
	\setlength{\tabcolsep}{0.5cm}
	\caption{Comparison of the magnetic parameters of [VO(HCOO)$_2\cdot$(H$_2$O)] with different reported spin-1/2 2D layered metal-organic compounds. Since $\theta_{\rm CW}$ for all the compounds is not available in the literature, for a quantitative comparison, we have tabulated both the $\theta_{\rm CW}/T_{\rm N}$ and $J_1/T_{\rm N}$ ratios.}
	\label{comp_J}
	\begin{tabular}{ccccccc}
		\hline \hline
		Compounds & $\theta_{\rm CW}$~(K) & $J_1/k_{\rm B}$~(K) & $T_{\rm N}$~(K) & $\theta_{CW}/T_{\rm N}$ & $J_1/T_{\rm N}$ & Refs. \\\hline
		[VO(HCOO)$_2\cdot$(H$_2$O)]  & 12 & 11  &  1.1 & 10.9 & 10 & this work \\
		Cu(PM)(EA)$_2$ & 3 & 6.8 & -- & -- &  --   &  [\onlinecite{Nath054409}] \\
		Cu(COOH)$_2$$ \cdot $4H$_2$O & 150 & 73 & 16.5 & 9.1 &  4.42  & [\onlinecite{Rnnow037202,Hanako541}] \\
		Cu(COOH)$_2$$ \cdot $2CO (NH$_2$)$_2$$ \cdot $2H$_2$O & -- & 70 & 15.5  & -- &  4.5  & [\onlinecite{Yamagata1179}] \\
		CuF$_2$$ \cdot $2H$_2$O & 37 & 26 & 10.9  & 3.39 & 2.38 & [\onlinecite{Abrahams56,Woodward144412}] \\\
		[Cu(C$ _5 $H$ _5 $NO)$_6$][BF$_4$]$_2$ & -- & 1.1 & 0.62 & -- &  1.77 & [\onlinecite{Algra24}] \\
		Cu(Pz)$_2$(ClO$_4$)$_2$ & 23.8 & 17.7 & 4.21  & 5.67 & 4.2 & [\onlinecite{Lancaster094421,Woodward4256}] \\\
		[Cu(Pz)$_2$(HF$_2$)]BF$_4$ & 8.1 & 2.85 & 1.54  & 5.25 & 1.8 & [\onlinecite{Manson4894}] \\
		(5MAP)$_2$CuBr$_4$ & -- & 6.5 & 3.8 & -- & 1.71 & [\onlinecite{Woodward144412}] \\
		(5CAP)$_2$CuBr$_4$ & -- & 8.5 & 5.08  & -- & 1.67 & [\onlinecite{Woodward144412}] \\
		(5CAP)$_2$CuCl$_4$ & -- & 1.14 & 0.74 & -- & 1.54 & [\onlinecite{Woodward144412}] \\
		(5MAP)$_2$CuCl$_4$ & -- & 0.76 & 0.44 & -- & 1.72 & [\onlinecite{Woodward144412}] \\
		\hline\hline
	\end{tabular}
\end{table*}
A most obvious feature of low-dimensional AFM spin systems is the occurrence of short-range order. Thus, the appearance of broad maximum in $\chi(T)$ and $C_{\rm p}(T)$ clearly suggests quasi-2D character of the compound.
Our experimental $\chi(T)$ data agree well with the HTSE for spin-$1/2$ 2D FSL model. Further confirmation about the two dimensionality can be obtained from the analysis of $C_{\rm mag}$. In low dimensional spin systems, the absolute value of $C_{\rm mag}$ at the maximum ($C_{\rm mag}^{\rm max}$) and the shape of the maximum are representative measures of dimensionality or quantum fluctuations and the temperature corresponding to the maximum ($T_{\rm C}^{\rm max}$) uniquely determines the exchange coupling.\cite{Nath024418,Bernu134409} With reduced dimensionality, the quantum fluctuations are enhanced which apparently suppresses the correlated spin excitations leading to a reduction in $C_{\rm mag}^{\rm max}$ and broadening of the maximum. For instance, in a non-frustrated 2D square lattice, a relatively higher value $C_{\rm mag}^{\rm max} \simeq 0.46 R = 3.82$~J/mol.K is expected at $T_{\rm C}^{\rm max}/J_{\rm 2D} = 0.60$.\cite{Makivic3562,Hofmann184502,Oitmaa14228} In case of uniform one-dimensional (1D) spin chains where the quantum fluctuations are more, this value decreases to $C_{\rm mag}^{\rm max} \simeq 0.35 R = 2.9$~J/mol.K at $T_{\rm C}^{\rm max}/J_{\rm 1D} = 0.48$ with a broad distribution.\cite{Bernu134409,Johnston9558} On the other hand, the triangular lattice (a frustrated 2D lattice) which is highly frustrated, shows even a lower ($C_{\rm mag}^{\rm max} \simeq 0.22 R = 1.83$~J/mol.K) value and broader maximum compared to uniform chains.\cite{Bernu134409} The effect of magnetic frustration not only suppresses the absolute value but also shifts the maximum towards lower temperatures. Clearly, our experimental value of $C_{\rm mag}^{\rm max} \simeq 3.6$~J/mol.K is more close to the one expected for the 2D square lattice model but much larger than the 1D model. This is a definite confirmation of the 2D character of the compound and a small reduction in the experimental $C_{\rm mag}$ value could be due to the bilayer nature of the spin-lattice and the effect of magnetic frustration.

To further justify the spin-lattice, QMC simulation was performed to simulate $M$ as a function of $H$ and compared with the experimental data at $T = 1.5$~K. The simulation was done assuming a pure 2D square lattice with no inter-layer couplings. For $J/k_{\rm B} = 10$~K, the simulation closely reproduces our experimental curve, especially in the low and high field regimes, implying that the spin-lattice is a 2D square lattice. However, a clear departure from the experimental data is noticed in the intermediate field range. This is because, our simulation is done assuming a purely non-frustrated square lattice model without any inter-layer couplings. As pointed out earlier, there are weak inter-layer couplings which are making a frustrated geometry. Thus, the curvature in $M$ vs $H$ curve could be attributed to this frustration effect and/or spatial anisotropy in the $ac$-plane which cannot be assessed from the magnetization data on the powder sample.\cite{Nath064422}

As shown in the inset of Fig.~\ref{Fig_7}, $T_{\rm N}$ shifts towards high temperatures almost linearly with increasing field in contrast to what is expected for a typical 3D AFM ordering. In several frustrated low-dimensional compounds it is observed that when magnetic field is applied, initially $T_{\rm N}$ moves slightly towards high temperatures. With further increase in field, it shifts back to low temperatures and reaches the fully polarized state.\cite{Nath064422,Nath024418,Nath214430} One possible explanation given is that the LRO in low-dimensional and/or frustrated spin systems is usually suppressed by quantum fluctuations. Magnetic field suppresses these fluctuations thereby enhancing $T_{\rm N}$ slightly at low fields (typically $\sim 1$~T). When the field is strong enough to overcome the AFM ordering, $T_{\rm N}$ is reduced thereby resulting in a belly shaped $H-T$ phase diagram.
However, in our compound $T_{\rm N}$ is still increasing even at a high field of 9~T which cannot be attributed to the effect of quantum fluctuations alone. Such an effect could also be due to the presence of large in-plane magnetic anisotropy, as expected from the crystal structure.\cite{Jongh2012}
Thus, the observed trend in the $H-T$ phase diagram is a signature of strong in-plane anisotropy in the present compound.

Moreover, in an AFM ordered state, one would expect a power law $(T^{\alpha}$) behaviour for $C_{\rm mag}(T)$ due to the spin wave excitations. For a 3D system, the exponent has a value $\alpha = 3$.\cite{Nath024431,Eisele929} On the other hand, for a 2D system, these excitations have a linear $k$ dependence around the Bragg points, leading to a $T^2$ dependence of $C_{\rm mag}$.\cite{Kitaoka3703} Similarly, for a 1D Heisenberg spin chain, $C_{\rm mag}(T)$ follows a linear behaviour with temperature, below $T_{\rm N}$. In our compound, $C_{\rm mag}(T)$ below $T_{\rm N}$ indeed follows a $T^\alpha$ behaviour with a reduced exponent $\alpha \simeq 2$. This observed quadratic $T$-dependence indicates the dominance of 2D AFM spin waves, below $T_{\rm N}$.\cite{Ramirez2070}
Finally, we made a comparison of our system with other reported compounds. Table~\ref{comp_J} summarizes the number of spin-$1/2$ metal-organic compounds with 2D geometry. Clearly, [VO(HCOO)$_2\cdot$(H$_2$O)] has the largest $\frac{\theta_{\rm CW}}{T_{\rm N}}$ or $\frac{J_1}{T_{\rm N}}$ ratio compared to other compounds making it the best example of a metal-organic based quasi-2D square lattice compound so far.

\section{\textbf{Conclusion}}
[VO(HCOO)$_2\cdot($H$_2$O)] is realized to be an exotic double layered square lattice compound. Indeed, the magnetic susceptibility, heat capacity, and high-field magnetization data could be described well by a spin-$1/2$ FSL model and consistently produce the intra-layer coupling $J_1/k_{\rm B} = (11 \pm 1)$~K. The saturation of exchange couplings happens at $H_{\rm S} \simeq 32$~T. It shows the onset of a magnetic LRO at a relatively low temperature $T_{\rm N}\simeq1.1$~K due to a weak inter-layer coupling. A much higher value of $\theta_{\rm CW}/T_{\rm N} \simeq 10.9$ or $J_1/T_{\rm N} \simeq 10$ compared to other 2D compounds makes it the best experimental realization of a quasi-2D square lattice so far among the metal-organic complexes. The $T^2$ dependence of $C_{\rm mag}$ at low temperatures further reflects the dominant role of 2D AFM magnons, below $T_{\rm N}$. Moreover, the disagreement of the high field magnetization data in the intermediate field range with that of the QMC simulation at low temperature can be attributed to the bilayered geometry and the effect of inter-layer frustration.

\section{\textbf{acknowledgement}}
UA, SG, and RN would like to acknowledge BRNS, India for financial support bearing sanction No.37(3)/14/26/2017-BRNS. AT also acknowledges the financial support from SERB, India for the grant EMR/2016/002637.


\begin{thebibliography}{57}%
	\makeatletter
	\providecommand \@ifxundefined [1]{%
		\@ifx{#1\undefined}
	}%
	\providecommand \@ifnum [1]{%
		\ifnum #1\expandafter \@firstoftwo
		\else \expandafter \@secondoftwo
		\fi
	}%
	\providecommand \@ifx [1]{%
		\ifx #1\expandafter \@firstoftwo
		\else \expandafter \@secondoftwo
		\fi
	}%
	\providecommand \natexlab [1]{#1}%
	\providecommand \enquote  [1]{``#1''}%
	\providecommand \bibnamefont  [1]{#1}%
	\providecommand \bibfnamefont [1]{#1}%
	\providecommand \citenamefont [1]{#1}%
	\providecommand \href@noop [0]{\@secondoftwo}%
	\providecommand \href [0]{\begingroup \@sanitize@url \@href}%
	\providecommand \@href[1]{\@@startlink{#1}\@@href}%
	\providecommand \@@href[1]{\endgroup#1\@@endlink}%
	\providecommand \@sanitize@url [0]{\catcode `\\12\catcode `\$12\catcode
		`\&12\catcode `\#12\catcode `\^12\catcode `\_12\catcode `\%12\relax}%
	\providecommand \@@startlink[1]{}%
	\providecommand \@@endlink[0]{}%
	\providecommand \url  [0]{\begingroup\@sanitize@url \@url }%
	\providecommand \@url [1]{\endgroup\@href {#1}{\urlprefix }}%
	\providecommand \urlprefix  [0]{URL }%
	\providecommand \Eprint [0]{\href }%
	\providecommand \doibase [0]{http://dx.doi.org/}%
	\providecommand \selectlanguage [0]{\@gobble}%
	\providecommand \bibinfo  [0]{\@secondoftwo}%
	\providecommand \bibfield  [0]{\@secondoftwo}%
	\providecommand \translation [1]{[#1]}%
	\providecommand \BibitemOpen [0]{}%
	\providecommand \bibitemStop [0]{}%
	\providecommand \bibitemNoStop [0]{.\EOS\space}%
	\providecommand \EOS [0]{\spacefactor3000\relax}%
	\providecommand \BibitemShut  [1]{\csname bibitem#1\endcsname}%
	\let\auto@bib@innerbib\@empty
	\bibitem [{\citenamefont {Manousakis}(1991)}]{Manousakis1}%
	\BibitemOpen
	\bibfield  {author} {\bibinfo {author} {\bibfnamefont {E.}~\bibnamefont
			{Manousakis}},\ }\href {\doibase 10.1103/RevModPhys.63.1} {\bibfield
		{journal} {\bibinfo  {journal} {Rev. Mod. Phys.}\ }\textbf {\bibinfo {volume}
			{63}},\ \bibinfo {pages} {1} (\bibinfo {year} {1991})}\BibitemShut {NoStop}%
	\bibitem [{\citenamefont {Makivi\ifmmode~\acute{c}\else \'{c}\fi{}}\ and\
		\citenamefont {Ding}(1991)}]{Makivic3562}%
	\BibitemOpen
	\bibfield  {author} {\bibinfo {author} {\bibfnamefont {M.~S.}\ \bibnamefont
			{Makivi\ifmmode~\acute{c}\else \'{c}\fi{}}}\ and\ \bibinfo {author}
		{\bibfnamefont {H.-Q.}\ \bibnamefont {Ding}},\ }\href {\doibase
		10.1103/PhysRevB.43.3562} {\bibfield  {journal} {\bibinfo  {journal} {Phys.
				Rev. B}\ }\textbf {\bibinfo {volume} {43}},\ \bibinfo {pages} {3562}
		(\bibinfo {year} {1991})}\BibitemShut {NoStop}%
	\bibitem [{\citenamefont {Kim}\ and\ \citenamefont {Troyer}(1998)}]{Kim2705}%
	\BibitemOpen
	\bibfield  {author} {\bibinfo {author} {\bibfnamefont {J.-K.}\ \bibnamefont
			{Kim}}\ and\ \bibinfo {author} {\bibfnamefont {M.}~\bibnamefont {Troyer}},\
	}\href {\doibase 10.1103/PhysRevLett.80.2705} {\bibfield  {journal} {\bibinfo
		{journal} {Phys. Rev. Lett.}\ }\textbf {\bibinfo {volume} {80}},\ \bibinfo
	{pages} {2705} (\bibinfo {year} {1998})}\BibitemShut {NoStop}%
\bibitem [{\citenamefont {Sandvik}(1997)}]{Sandvik11678}%
\BibitemOpen
\bibfield  {author} {\bibinfo {author} {\bibfnamefont {A.~W.}\ \bibnamefont
		{Sandvik}},\ }\href {\doibase 10.1103/PhysRevB.56.11678} {\bibfield
	{journal} {\bibinfo  {journal} {Phys. Rev. B}\ }\textbf {\bibinfo {volume}
		{56}},\ \bibinfo {pages} {11678} (\bibinfo {year} {1997})}\BibitemShut
{NoStop}%
\bibitem [{\citenamefont {Mermin}\ and\ \citenamefont
	{Wagner}(1966)}]{Mermin1133}%
\BibitemOpen
\bibfield  {author} {\bibinfo {author} {\bibfnamefont {N.~D.}\ \bibnamefont
		{Mermin}}\ and\ \bibinfo {author} {\bibfnamefont {H.}~\bibnamefont
		{Wagner}},\ }\href {\doibase 10.1103/PhysRevLett.17.1133} {\bibfield
	{journal} {\bibinfo  {journal} {Phys. Rev. Lett.}\ }\textbf {\bibinfo
		{volume} {17}},\ \bibinfo {pages} {1133} (\bibinfo {year}
	{1966})}\BibitemShut {NoStop}%
\bibitem [{\citenamefont {Yasuda}\ \emph {et~al.}(2005)\citenamefont {Yasuda},
	\citenamefont {Todo}, \citenamefont {Hukushima}, \citenamefont {Alet},
	\citenamefont {Keller}, \citenamefont {Troyer},\ and\ \citenamefont
	{Takayama}}]{Yasuda217201}%
\BibitemOpen
\bibfield  {author} {\bibinfo {author} {\bibfnamefont {C.}~\bibnamefont
		{Yasuda}}, \bibinfo {author} {\bibfnamefont {S.}~\bibnamefont {Todo}},
	\bibinfo {author} {\bibfnamefont {K.}~\bibnamefont {Hukushima}}, \bibinfo
	{author} {\bibfnamefont {F.}~\bibnamefont {Alet}}, \bibinfo {author}
	{\bibfnamefont {M.}~\bibnamefont {Keller}}, \bibinfo {author} {\bibfnamefont
		{M.}~\bibnamefont {Troyer}}, \ and\ \bibinfo {author} {\bibfnamefont
		{H.}~\bibnamefont {Takayama}},\ }\href {\doibase
	10.1103/PhysRevLett.94.217201} {\bibfield  {journal} {\bibinfo  {journal}
		{Phys. Rev. Lett.}\ }\textbf {\bibinfo {volume} {94}},\ \bibinfo {pages}
	{217201} (\bibinfo {year} {2005})}\BibitemShut {NoStop}%
\bibitem [{\citenamefont {{Shannon, N.}}\ \emph {et~al.}(2004)\citenamefont
	{{Shannon, N.}}, \citenamefont {{Schmidt, B.}}, \citenamefont {{Penc, K.}},\
	and\ \citenamefont {{Thalmeier, P.}}}]{Shannon599}%
\BibitemOpen
\bibfield  {author} {\bibinfo {author} {\bibnamefont {{Shannon, N.}}},
	\bibinfo {author} {\bibnamefont {{Schmidt, B.}}}, \bibinfo {author}
	{\bibnamefont {{Penc, K.}}}, \ and\ \bibinfo {author} {\bibnamefont
		{{Thalmeier, P.}}},\ }\href {\doibase 10.1140/epjb/e2004-00156-3} {\bibfield
	{journal} {\bibinfo  {journal} {Eur. Phys. J. B}\ }\textbf {\bibinfo {volume}
		{38}},\ \bibinfo {pages} {599} (\bibinfo {year} {2004})}\BibitemShut
{NoStop}%
\bibitem [{\citenamefont {Shannon}\ \emph {et~al.}(2006)\citenamefont
	{Shannon}, \citenamefont {Momoi},\ and\ \citenamefont
	{Sindzingre}}]{Shannon027213}%
\BibitemOpen
\bibfield  {author} {\bibinfo {author} {\bibfnamefont {N.}~\bibnamefont
		{Shannon}}, \bibinfo {author} {\bibfnamefont {T.}~\bibnamefont {Momoi}}, \
	and\ \bibinfo {author} {\bibfnamefont {P.}~\bibnamefont {Sindzingre}},\
}\href {\doibase 10.1103/PhysRevLett.96.027213} {\bibfield  {journal}
{\bibinfo  {journal} {Phys. Rev. Lett.}\ }\textbf {\bibinfo {volume} {96}},\
\bibinfo {pages} {027213} (\bibinfo {year} {2006})}\BibitemShut {NoStop}%
\bibitem [{\citenamefont {Nath}\ \emph
	{et~al.}(2008{\natexlab{a}})\citenamefont {Nath}, \citenamefont {Tsirlin},
	\citenamefont {Rosner},\ and\ \citenamefont {Geibel}}]{Nath064422}%
\BibitemOpen
\bibfield  {author} {\bibinfo {author} {\bibfnamefont {R.}~\bibnamefont
		{Nath}}, \bibinfo {author} {\bibfnamefont {A.~A.}\ \bibnamefont {Tsirlin}},
	\bibinfo {author} {\bibfnamefont {H.}~\bibnamefont {Rosner}}, \ and\ \bibinfo
	{author} {\bibfnamefont {C.}~\bibnamefont {Geibel}},\ }\href {\doibase
	10.1103/PhysRevB.78.064422} {\bibfield  {journal} {\bibinfo  {journal} {Phys.
			Rev. B}\ }\textbf {\bibinfo {volume} {78}},\ \bibinfo {pages} {064422}
	(\bibinfo {year} {2008}{\natexlab{a}})}\BibitemShut {NoStop}%
\bibitem [{\citenamefont {Nath}\ \emph {et~al.}(2009)\citenamefont {Nath},
	\citenamefont {Furukawa}, \citenamefont {Borsa}, \citenamefont {Kaul},
	\citenamefont {Baenitz}, \citenamefont {Geibel},\ and\ \citenamefont
	{Johnston}}]{Nath214430}%
\BibitemOpen
\bibfield  {author} {\bibinfo {author} {\bibfnamefont {R.}~\bibnamefont
		{Nath}}, \bibinfo {author} {\bibfnamefont {Y.}~\bibnamefont {Furukawa}},
	\bibinfo {author} {\bibfnamefont {F.}~\bibnamefont {Borsa}}, \bibinfo
	{author} {\bibfnamefont {E.~E.}\ \bibnamefont {Kaul}}, \bibinfo {author}
	{\bibfnamefont {M.}~\bibnamefont {Baenitz}}, \bibinfo {author} {\bibfnamefont
		{C.}~\bibnamefont {Geibel}}, \ and\ \bibinfo {author} {\bibfnamefont {D.~C.}\
		\bibnamefont {Johnston}},\ }\href {\doibase 10.1103/PhysRevB.80.214430}
{\bibfield  {journal} {\bibinfo  {journal} {Phys. Rev. B}\ }\textbf {\bibinfo
		{volume} {80}},\ \bibinfo {pages} {214430} (\bibinfo {year}
	{2009})}\BibitemShut {NoStop}%
\bibitem [{\citenamefont {Tsyrulin}\ \emph {et~al.}(2009)\citenamefont
	{Tsyrulin}, \citenamefont {Pardini}, \citenamefont {Singh}, \citenamefont
	{Xiao}, \citenamefont {Link}, \citenamefont {Schneidewind}, \citenamefont
	{Hiess}, \citenamefont {Landee}, \citenamefont {Turnbull},\ and\
	\citenamefont {Kenzelmann}}]{Tsyrulin197201}%
\BibitemOpen
\bibfield  {author} {\bibinfo {author} {\bibfnamefont {N.}~\bibnamefont
		{Tsyrulin}}, \bibinfo {author} {\bibfnamefont {T.}~\bibnamefont {Pardini}},
	\bibinfo {author} {\bibfnamefont {R.~R.~P.}\ \bibnamefont {Singh}}, \bibinfo
	{author} {\bibfnamefont {F.}~\bibnamefont {Xiao}}, \bibinfo {author}
	{\bibfnamefont {P.}~\bibnamefont {Link}}, \bibinfo {author} {\bibfnamefont
		{A.}~\bibnamefont {Schneidewind}}, \bibinfo {author} {\bibfnamefont
		{A.}~\bibnamefont {Hiess}}, \bibinfo {author} {\bibfnamefont {C.~P.}\
		\bibnamefont {Landee}}, \bibinfo {author} {\bibfnamefont {M.~M.}\
		\bibnamefont {Turnbull}}, \ and\ \bibinfo {author} {\bibfnamefont
		{M.}~\bibnamefont {Kenzelmann}},\ }\href {\doibase
	10.1103/PhysRevLett.102.197201} {\bibfield  {journal} {\bibinfo  {journal}
		{Phys. Rev. Lett.}\ }\textbf {\bibinfo {volume} {102}},\ \bibinfo {pages}
	{197201} (\bibinfo {year} {2009})}\BibitemShut {NoStop}%
\bibitem [{\citenamefont {Lee}(2007)}]{Lee012501}%
\BibitemOpen
\bibfield  {author} {\bibinfo {author} {\bibfnamefont {P.~A.}\ \bibnamefont
		{Lee}},\ }\href {\doibase 10.1088/0034-4885/71/1/012501} {\bibfield
	{journal} {\bibinfo  {journal} {Rep. Prog. Phys.}\ }\textbf {\bibinfo
		{volume} {71}},\ \bibinfo {pages} {012501} (\bibinfo {year}
	{2007})}\BibitemShut {NoStop}%
\bibitem [{\citenamefont {Jain}\ \emph {et~al.}(2017)\citenamefont {Jain},
	\citenamefont {Krautloher}, \citenamefont {Porras}, \citenamefont {Ryu},
	\citenamefont {Chen}, \citenamefont {Abernathy}, \citenamefont {Park},
	\citenamefont {Ivanov}, \citenamefont {Chaloupka}, \citenamefont
	{Khaliullin}, \citenamefont {Keimer},\ and\ \citenamefont {Kim}}]{Jain2017}%
\BibitemOpen
\bibfield  {author} {\bibinfo {author} {\bibfnamefont {A.}~\bibnamefont
		{Jain}}, \bibinfo {author} {\bibfnamefont {M.}~\bibnamefont {Krautloher}},
	\bibinfo {author} {\bibfnamefont {J.}~\bibnamefont {Porras}}, \bibinfo
	{author} {\bibfnamefont {G.}~\bibnamefont {Ryu}}, \bibinfo {author}
	{\bibfnamefont {D.}~\bibnamefont {Chen}}, \bibinfo {author} {\bibfnamefont
		{D.}~\bibnamefont {Abernathy}}, \bibinfo {author} {\bibfnamefont
		{J.}~\bibnamefont {Park}}, \bibinfo {author} {\bibfnamefont {A.}~\bibnamefont
		{Ivanov}}, \bibinfo {author} {\bibfnamefont {J.}~\bibnamefont {Chaloupka}},
	\bibinfo {author} {\bibfnamefont {G.}~\bibnamefont {Khaliullin}}, \bibinfo
	{author} {\bibfnamefont {B.}~\bibnamefont {Keimer}}, \ and\ \bibinfo {author}
	{\bibfnamefont {B.}~\bibnamefont {Kim}},\ }\href@noop {} {\bibfield
	{journal} {\bibinfo  {journal} {Nat. Phys.}\ }\textbf {\bibinfo {volume}
		{13}},\ \bibinfo {pages} {633} (\bibinfo {year} {2017})}\BibitemShut
{NoStop}%
\bibitem [{\citenamefont {Pekker}\ and\ \citenamefont
	{Varma}(2015)}]{Pekker269}%
\BibitemOpen
\bibfield  {author} {\bibinfo {author} {\bibfnamefont {D.}~\bibnamefont
		{Pekker}}\ and\ \bibinfo {author} {\bibfnamefont {C.}~\bibnamefont {Varma}},\
}\href@noop {} {\bibfield  {journal} {\bibinfo  {journal} {Annu. Rev.
		Condens. Matter Phys.}\ }\textbf {\bibinfo {volume} {6}},\ \bibinfo {pages}
{269} (\bibinfo {year} {2015})}\BibitemShut {NoStop}%
\bibitem [{\citenamefont {Goddard}\ \emph {et~al.}(2012)\citenamefont
	{Goddard}, \citenamefont {Manson}, \citenamefont {Singleton}, \citenamefont
	{Franke}, \citenamefont {Lancaster}, \citenamefont {Steele}, \citenamefont
	{Blundell}, \citenamefont {Baines}, \citenamefont {Pratt}, \citenamefont
	{McDonald}, \citenamefont {Ayala-Valenzuela}, \citenamefont {Corbey},
	\citenamefont {Southerland}, \citenamefont {Sengupta},\ and\ \citenamefont
	{Schlueter}}]{Goddard077208}%
\BibitemOpen
\bibfield  {author} {\bibinfo {author} {\bibfnamefont {P.~A.}\ \bibnamefont
		{Goddard}}, \bibinfo {author} {\bibfnamefont {J.~L.}\ \bibnamefont {Manson}},
	\bibinfo {author} {\bibfnamefont {J.}~\bibnamefont {Singleton}}, \bibinfo
	{author} {\bibfnamefont {I.}~\bibnamefont {Franke}}, \bibinfo {author}
	{\bibfnamefont {T.}~\bibnamefont {Lancaster}}, \bibinfo {author}
	{\bibfnamefont {A.~J.}\ \bibnamefont {Steele}}, \bibinfo {author}
	{\bibfnamefont {S.~J.}\ \bibnamefont {Blundell}}, \bibinfo {author}
	{\bibfnamefont {C.}~\bibnamefont {Baines}}, \bibinfo {author} {\bibfnamefont
		{F.~L.}\ \bibnamefont {Pratt}}, \bibinfo {author} {\bibfnamefont {R.~D.}\
		\bibnamefont {McDonald}}, \bibinfo {author} {\bibfnamefont {O.~E.}\
		\bibnamefont {Ayala-Valenzuela}}, \bibinfo {author} {\bibfnamefont {J.~F.}\
		\bibnamefont {Corbey}}, \bibinfo {author} {\bibfnamefont {H.~I.}\
		\bibnamefont {Southerland}}, \bibinfo {author} {\bibfnamefont
		{P.}~\bibnamefont {Sengupta}}, \ and\ \bibinfo {author} {\bibfnamefont
		{J.~A.}\ \bibnamefont {Schlueter}},\ }\href {\doibase
	10.1103/PhysRevLett.108.077208} {\bibfield  {journal} {\bibinfo  {journal}
		{Phys. Rev. Lett.}\ }\textbf {\bibinfo {volume} {108}},\ \bibinfo {pages}
	{077208} (\bibinfo {year} {2012})}\BibitemShut {NoStop}%
\bibitem [{\citenamefont {Nath}\ \emph {et~al.}(2015)\citenamefont {Nath},
	\citenamefont {Padmanabhan}, \citenamefont {Baby}, \citenamefont
	{Thirumurugan}, \citenamefont {Ehlers}, \citenamefont {Hemmida},
	\citenamefont {Krug~von Nidda},\ and\ \citenamefont {Tsirlin}}]{Nath054409}%
\BibitemOpen
\bibfield  {author} {\bibinfo {author} {\bibfnamefont {R.}~\bibnamefont
		{Nath}}, \bibinfo {author} {\bibfnamefont {M.}~\bibnamefont {Padmanabhan}},
	\bibinfo {author} {\bibfnamefont {S.}~\bibnamefont {Baby}}, \bibinfo {author}
	{\bibfnamefont {A.}~\bibnamefont {Thirumurugan}}, \bibinfo {author}
	{\bibfnamefont {D.}~\bibnamefont {Ehlers}}, \bibinfo {author} {\bibfnamefont
		{M.}~\bibnamefont {Hemmida}}, \bibinfo {author} {\bibfnamefont {H.-A.}\
		\bibnamefont {Krug~von Nidda}}, \ and\ \bibinfo {author} {\bibfnamefont
		{A.~A.}\ \bibnamefont {Tsirlin}},\ }\href {\doibase
	10.1103/PhysRevB.91.054409} {\bibfield  {journal} {\bibinfo  {journal} {Phys.
			Rev. B}\ }\textbf {\bibinfo {volume} {91}},\ \bibinfo {pages} {054409}
	(\bibinfo {year} {2015})}\BibitemShut {NoStop}%
\bibitem [{\citenamefont {R\o{}nnow}\ \emph {et~al.}(2001)\citenamefont
	{R\o{}nnow}, \citenamefont {McMorrow}, \citenamefont {Coldea}, \citenamefont
	{Harrison}, \citenamefont {Youngson}, \citenamefont {Perring}, \citenamefont
	{Aeppli}, \citenamefont {Sylju\aa{}sen}, \citenamefont {Lefmann},\ and\
	\citenamefont {Rischel}}]{Rnnow037202}%
\BibitemOpen
\bibfield  {author} {\bibinfo {author} {\bibfnamefont {H.~M.}\ \bibnamefont
		{R\o{}nnow}}, \bibinfo {author} {\bibfnamefont {D.~F.}\ \bibnamefont
		{McMorrow}}, \bibinfo {author} {\bibfnamefont {R.}~\bibnamefont {Coldea}},
	\bibinfo {author} {\bibfnamefont {A.}~\bibnamefont {Harrison}}, \bibinfo
	{author} {\bibfnamefont {I.~D.}\ \bibnamefont {Youngson}}, \bibinfo {author}
	{\bibfnamefont {T.~G.}\ \bibnamefont {Perring}}, \bibinfo {author}
	{\bibfnamefont {G.}~\bibnamefont {Aeppli}}, \bibinfo {author} {\bibfnamefont
		{O.}~\bibnamefont {Sylju\aa{}sen}}, \bibinfo {author} {\bibfnamefont
		{K.}~\bibnamefont {Lefmann}}, \ and\ \bibinfo {author} {\bibfnamefont
		{C.}~\bibnamefont {Rischel}},\ }\href {\doibase
	10.1103/PhysRevLett.87.037202} {\bibfield  {journal} {\bibinfo  {journal}
		{Phys. Rev. Lett.}\ }\textbf {\bibinfo {volume} {87}},\ \bibinfo {pages}
	{037202} (\bibinfo {year} {2001})}\BibitemShut {NoStop}%
\bibitem [{\citenamefont {Tsyrulin}\ \emph {et~al.}(2010)\citenamefont
	{Tsyrulin}, \citenamefont {Xiao}, \citenamefont {Schneidewind}, \citenamefont
	{Link}, \citenamefont {R\o{}nnow}, \citenamefont {Gavilano}, \citenamefont
	{Landee}, \citenamefont {Turnbull},\ and\ \citenamefont
	{Kenzelmann}}]{Tsyrulin134409}%
\BibitemOpen
\bibfield  {author} {\bibinfo {author} {\bibfnamefont {N.}~\bibnamefont
		{Tsyrulin}}, \bibinfo {author} {\bibfnamefont {F.}~\bibnamefont {Xiao}},
	\bibinfo {author} {\bibfnamefont {A.}~\bibnamefont {Schneidewind}}, \bibinfo
	{author} {\bibfnamefont {P.}~\bibnamefont {Link}}, \bibinfo {author}
	{\bibfnamefont {H.~M.}\ \bibnamefont {R\o{}nnow}}, \bibinfo {author}
	{\bibfnamefont {J.}~\bibnamefont {Gavilano}}, \bibinfo {author}
	{\bibfnamefont {C.~P.}\ \bibnamefont {Landee}}, \bibinfo {author}
	{\bibfnamefont {M.~M.}\ \bibnamefont {Turnbull}}, \ and\ \bibinfo {author}
	{\bibfnamefont {M.}~\bibnamefont {Kenzelmann}},\ }\href {\doibase
	10.1103/PhysRevB.81.134409} {\bibfield  {journal} {\bibinfo  {journal} {Phys.
			Rev. B}\ }\textbf {\bibinfo {volume} {81}},\ \bibinfo {pages} {134409}
	(\bibinfo {year} {2010})}\BibitemShut {NoStop}%
\bibitem [{\citenamefont {Siahatgar}\ \emph {et~al.}(2011)\citenamefont
	{Siahatgar}, \citenamefont {Schmidt},\ and\ \citenamefont
	{Thalmeier}}]{Siahatgar064431}%
\BibitemOpen
\bibfield  {author} {\bibinfo {author} {\bibfnamefont {M.}~\bibnamefont
		{Siahatgar}}, \bibinfo {author} {\bibfnamefont {B.}~\bibnamefont {Schmidt}},
	\ and\ \bibinfo {author} {\bibfnamefont {P.}~\bibnamefont {Thalmeier}},\
}\href {\doibase 10.1103/PhysRevB.84.064431} {\bibfield  {journal} {\bibinfo
	{journal} {Phys. Rev. B}\ }\textbf {\bibinfo {volume} {84}},\ \bibinfo
{pages} {064431} (\bibinfo {year} {2011})}\BibitemShut {NoStop}%
\bibitem [{\citenamefont {Lancaster}\ \emph {et~al.}(2007)\citenamefont
	{Lancaster}, \citenamefont {Blundell}, \citenamefont {Brooks}, \citenamefont
	{Baker}, \citenamefont {Pratt}, \citenamefont {Manson}, \citenamefont
	{Conner}, \citenamefont {Xiao}, \citenamefont {Landee}, \citenamefont
	{Chaves}, \citenamefont {Soriano}, \citenamefont {Novak}, \citenamefont
	{Papageorgiou}, \citenamefont {Bianchi}, \citenamefont {Herrmannsd\"orfer},
	\citenamefont {Wosnitza},\ and\ \citenamefont {Schlueter}}]{Lancaster094421}%
\BibitemOpen
\bibfield  {author} {\bibinfo {author} {\bibfnamefont {T.}~\bibnamefont
		{Lancaster}}, \bibinfo {author} {\bibfnamefont {S.~J.}\ \bibnamefont
		{Blundell}}, \bibinfo {author} {\bibfnamefont {M.~L.}\ \bibnamefont
		{Brooks}}, \bibinfo {author} {\bibfnamefont {P.~J.}\ \bibnamefont {Baker}},
	\bibinfo {author} {\bibfnamefont {F.~L.}\ \bibnamefont {Pratt}}, \bibinfo
	{author} {\bibfnamefont {J.~L.}\ \bibnamefont {Manson}}, \bibinfo {author}
	{\bibfnamefont {M.~M.}\ \bibnamefont {Conner}}, \bibinfo {author}
	{\bibfnamefont {F.}~\bibnamefont {Xiao}}, \bibinfo {author} {\bibfnamefont
		{C.~P.}\ \bibnamefont {Landee}}, \bibinfo {author} {\bibfnamefont {F.~A.}\
		\bibnamefont {Chaves}}, \bibinfo {author} {\bibfnamefont {S.}~\bibnamefont
		{Soriano}}, \bibinfo {author} {\bibfnamefont {M.~A.}\ \bibnamefont {Novak}},
	\bibinfo {author} {\bibfnamefont {T.~P.}\ \bibnamefont {Papageorgiou}},
	\bibinfo {author} {\bibfnamefont {A.~D.}\ \bibnamefont {Bianchi}}, \bibinfo
	{author} {\bibfnamefont {T.}~\bibnamefont {Herrmannsd\"orfer}}, \bibinfo
	{author} {\bibfnamefont {J.}~\bibnamefont {Wosnitza}}, \ and\ \bibinfo
	{author} {\bibfnamefont {J.~A.}\ \bibnamefont {Schlueter}},\ }\href {\doibase
	10.1103/PhysRevB.75.094421} {\bibfield  {journal} {\bibinfo  {journal} {Phys.
			Rev. B}\ }\textbf {\bibinfo {volume} {75}},\ \bibinfo {pages} {094421}
	(\bibinfo {year} {2007})}\BibitemShut {NoStop}%
\bibitem [{\citenamefont {Lefebvre}\ \emph {et~al.}(2000)\citenamefont
	{Lefebvre}, \citenamefont {Wzietek}, \citenamefont {Brown}, \citenamefont
	{Bourbonnais}, \citenamefont {J\'erome}, \citenamefont {M\'ezi\`ere},
	\citenamefont {Fourmigu\'e},\ and\ \citenamefont {Batail}}]{Lefebvre5420}%
\BibitemOpen
\bibfield  {author} {\bibinfo {author} {\bibfnamefont {S.}~\bibnamefont
		{Lefebvre}}, \bibinfo {author} {\bibfnamefont {P.}~\bibnamefont {Wzietek}},
	\bibinfo {author} {\bibfnamefont {S.}~\bibnamefont {Brown}}, \bibinfo
	{author} {\bibfnamefont {C.}~\bibnamefont {Bourbonnais}}, \bibinfo {author}
	{\bibfnamefont {D.}~\bibnamefont {J\'erome}}, \bibinfo {author}
	{\bibfnamefont {C.}~\bibnamefont {M\'ezi\`ere}}, \bibinfo {author}
	{\bibfnamefont {M.}~\bibnamefont {Fourmigu\'e}}, \ and\ \bibinfo {author}
	{\bibfnamefont {P.}~\bibnamefont {Batail}},\ }\href {\doibase
	10.1103/PhysRevLett.85.5420} {\bibfield  {journal} {\bibinfo  {journal}
		{Phys. Rev. Lett.}\ }\textbf {\bibinfo {volume} {85}},\ \bibinfo {pages}
	{5420} (\bibinfo {year} {2000})}\BibitemShut {NoStop}%
\bibitem [{\citenamefont {Nam}\ \emph {et~al.}(2007)\citenamefont {Nam},
	\citenamefont {Ardavan}, \citenamefont {Blundell},\ and\ \citenamefont
	{Schlueter}}]{Nam584}%
\BibitemOpen
\bibfield  {author} {\bibinfo {author} {\bibfnamefont {M.-S.}\ \bibnamefont
		{Nam}}, \bibinfo {author} {\bibfnamefont {A.}~\bibnamefont {Ardavan}},
	\bibinfo {author} {\bibfnamefont {S.~J.}\ \bibnamefont {Blundell}}, \ and\
	\bibinfo {author} {\bibfnamefont {J.~A.}\ \bibnamefont {Schlueter}},\
}\href@noop {} {\bibfield  {journal} {\bibinfo  {journal} {Nature}\ }\textbf
{\bibinfo {volume} {449}},\ \bibinfo {pages} {584} (\bibinfo {year}
{2007})}\BibitemShut {NoStop}%
\bibitem [{\citenamefont {T.~Ishiguro}\ and\ \citenamefont
	{G.Saito}(2006)}]{Ishiguro2006}%
\BibitemOpen
\bibfield  {author} {\bibinfo {author} {\bibfnamefont {K.~Y.}\ \bibnamefont
		{T.~Ishiguro}}\ and\ \bibinfo {author} {\bibnamefont {G.Saito}},\ }\href@noop
{} {\emph {\bibinfo {title} {Organic Superconductors}}},\ \bibinfo {edition}
{2nd}\ ed.\ (\bibinfo  {publisher} {Springer},\ \bibinfo {address} {Berlin},\
\bibinfo {year} {2006})\BibitemShut {NoStop}%
\bibitem [{\citenamefont {Manna}\ \emph {et~al.}(2010)\citenamefont {Manna},
	\citenamefont {de~Souza}, \citenamefont {Br\"uhl}, \citenamefont
	{Schlueter},\ and\ \citenamefont {Lang}}]{Manna016403}%
\BibitemOpen
\bibfield  {author} {\bibinfo {author} {\bibfnamefont {R.~S.}\ \bibnamefont
		{Manna}}, \bibinfo {author} {\bibfnamefont {M.}~\bibnamefont {de~Souza}},
	\bibinfo {author} {\bibfnamefont {A.}~\bibnamefont {Br\"uhl}}, \bibinfo
	{author} {\bibfnamefont {J.~A.}\ \bibnamefont {Schlueter}}, \ and\ \bibinfo
	{author} {\bibfnamefont {M.}~\bibnamefont {Lang}},\ }\href {\doibase
	10.1103/PhysRevLett.104.016403} {\bibfield  {journal} {\bibinfo  {journal}
		{Phys. Rev. Lett.}\ }\textbf {\bibinfo {volume} {104}},\ \bibinfo {pages}
	{016403} (\bibinfo {year} {2010})}\BibitemShut {NoStop}%
\bibitem [{\citenamefont {Mootz}\ and\ \citenamefont
	{Seidel}(1987)}]{Mootz1218}%
\BibitemOpen
\bibfield  {author} {\bibinfo {author} {\bibfnamefont {D.}~\bibnamefont
		{Mootz}}\ and\ \bibinfo {author} {\bibfnamefont {R.}~\bibnamefont {Seidel}},\
}\href@noop {} {\bibfield  {journal} {\bibinfo  {journal} {Acta Crystallogr.,
		Sect. C}\ }\textbf {\bibinfo {volume} {43}},\ \bibinfo {pages} {1218}
(\bibinfo {year} {1987})}\BibitemShut {NoStop}%
\bibitem [{\citenamefont {Gilson}(1995)}]{Gilson136}%
\BibitemOpen
\bibfield  {author} {\bibinfo {author} {\bibfnamefont {T.~R.}\ \bibnamefont
		{Gilson}},\ }\href {\doibase http://dx.doi.org/10.1006/jssc.1995.1256}
{\bibfield  {journal} {\bibinfo  {journal} {J. Solid State Chem}\ }\textbf
	{\bibinfo {volume} {117}},\ \bibinfo {pages} {136 } (\bibinfo {year}
	{1995})}\BibitemShut {NoStop}%
\bibitem [{\citenamefont {Rodríguez-Carvajal}(1993)}]{Carvajal55}%
\BibitemOpen
\bibfield  {author} {\bibinfo {author} {\bibfnamefont {J.}~\bibnamefont
		{Rodríguez-Carvajal}},\ }\href {\doibase
	http://dx.doi.org/10.1016/0921-4526(93)90108-I} {\bibfield  {journal}
	{\bibinfo  {journal} {Physica B: Condens. Matter}\ }\textbf {\bibinfo
		{volume} {192}},\ \bibinfo {pages} {55 } (\bibinfo {year}
	{1993})}\BibitemShut {NoStop}%
\bibitem [{ALP()}]{ALPS}%
\BibitemOpen
\href@noop {} {\enquote {\bibinfo {title} {{ALPS} project},}\ }\bibinfo
{howpublished} {\url{http://alps.comp-phys.org/}}\BibitemShut {NoStop}%
\bibitem [{\citenamefont {Sandvik}(1999)}]{SandvikR14157}%
\BibitemOpen
\bibfield  {author} {\bibinfo {author} {\bibfnamefont {A.~W.}\ \bibnamefont
		{Sandvik}},\ }\href {\doibase 10.1103/PhysRevB.59.R14157} {\bibfield
	{journal} {\bibinfo  {journal} {Phys. Rev. B}\ }\textbf {\bibinfo {volume}
		{59}},\ \bibinfo {pages} {R14157} (\bibinfo {year} {1999})}\BibitemShut
{NoStop}%
\bibitem [{\citenamefont {Alet}\ \emph {et~al.}(2005)\citenamefont {Alet},
	\citenamefont {Wessel},\ and\ \citenamefont {Troyer}}]{Alet036706}%
\BibitemOpen
\bibfield  {author} {\bibinfo {author} {\bibfnamefont {F.}~\bibnamefont
		{Alet}}, \bibinfo {author} {\bibfnamefont {S.}~\bibnamefont {Wessel}}, \ and\
	\bibinfo {author} {\bibfnamefont {M.}~\bibnamefont {Troyer}},\ }\href
{\doibase 10.1103/PhysRevE.71.036706} {\bibfield  {journal} {\bibinfo
		{journal} {Phys. Rev. E}\ }\textbf {\bibinfo {volume} {71}},\ \bibinfo
	{pages} {036706} (\bibinfo {year} {2005})}\BibitemShut {NoStop}%
\bibitem [{\citenamefont {Pollet}\ \emph {et~al.}(2004)\citenamefont {Pollet},
	\citenamefont {Rombouts}, \citenamefont {Van~Houcke},\ and\ \citenamefont
	{Heyde}}]{Pollet056705}%
\BibitemOpen
\bibfield  {author} {\bibinfo {author} {\bibfnamefont {L.}~\bibnamefont
		{Pollet}}, \bibinfo {author} {\bibfnamefont {S.~M.~A.}\ \bibnamefont
		{Rombouts}}, \bibinfo {author} {\bibfnamefont {K.}~\bibnamefont
		{Van~Houcke}}, \ and\ \bibinfo {author} {\bibfnamefont {K.}~\bibnamefont
		{Heyde}},\ }\href {\doibase 10.1103/PhysRevE.70.056705} {\bibfield  {journal}
	{\bibinfo  {journal} {Phys. Rev. E}\ }\textbf {\bibinfo {volume} {70}},\
	\bibinfo {pages} {056705} (\bibinfo {year} {2004})}\BibitemShut {NoStop}%
\bibitem [{\citenamefont {Domb}\ and\ \citenamefont
	{Miedema}(1964)}]{Domb1964}%
\BibitemOpen
\bibfield  {author} {\bibinfo {author} {\bibfnamefont {C.}~\bibnamefont
		{Domb}}\ and\ \bibinfo {author} {\bibfnamefont {A.~R.}\ \bibnamefont
		{Miedema}},\ }\href@noop {} {\emph {\bibinfo {title} {Progress in Low
			Temperature Physics}}},\ edited by\ \bibinfo {editor} {\bibfnamefont {C.~J.}\
	\bibnamefont {Gorter}},\ Vol.~\bibinfo {volume} {4}\ (\bibinfo  {publisher}
{North Holland},\ \bibinfo {address} {Amsterdam},\ \bibinfo {year}
{1964})\BibitemShut {NoStop}%
\bibitem [{\citenamefont {Rosner}\ \emph {et~al.}(2003)\citenamefont {Rosner},
	\citenamefont {Singh}, \citenamefont {Zheng}, \citenamefont {Oitmaa},\ and\
	\citenamefont {Pickett}}]{Rosner014416}%
\BibitemOpen
\bibfield  {author} {\bibinfo {author} {\bibfnamefont {H.}~\bibnamefont
		{Rosner}}, \bibinfo {author} {\bibfnamefont {R.~R.~P.}\ \bibnamefont
		{Singh}}, \bibinfo {author} {\bibfnamefont {W.~H.}\ \bibnamefont {Zheng}},
	\bibinfo {author} {\bibfnamefont {J.}~\bibnamefont {Oitmaa}}, \ and\ \bibinfo
	{author} {\bibfnamefont {W.~E.}\ \bibnamefont {Pickett}},\ }\href {\doibase
	10.1103/PhysRevB.67.014416} {\bibfield  {journal} {\bibinfo  {journal} {Phys.
			Rev. B}\ }\textbf {\bibinfo {volume} {67}},\ \bibinfo {pages} {014416}
	(\bibinfo {year} {2003})}\BibitemShut {NoStop}%
\bibitem [{\citenamefont {Schmidt}\ \emph {et~al.}(2011)\citenamefont
	{Schmidt}, \citenamefont {Lohmann},\ and\ \citenamefont
	{Richter}}]{Schmidt104443}%
\BibitemOpen
\bibfield  {author} {\bibinfo {author} {\bibfnamefont {H.-J.}\ \bibnamefont
		{Schmidt}}, \bibinfo {author} {\bibfnamefont {A.}~\bibnamefont {Lohmann}}, \
	and\ \bibinfo {author} {\bibfnamefont {J.}~\bibnamefont {Richter}},\ }\href
{\doibase 10.1103/PhysRevB.84.104443} {\bibfield  {journal} {\bibinfo
		{journal} {Phys. Rev. B}\ }\textbf {\bibinfo {volume} {84}},\ \bibinfo
	{pages} {104443} (\bibinfo {year} {2011})}\BibitemShut {NoStop}%
\bibitem [{\citenamefont {Majlis}\ \emph {et~al.}(1992)\citenamefont {Majlis},
	\citenamefont {Selzer},\ and\ \citenamefont {Strinati}}]{Majlis7872}%
\BibitemOpen
\bibfield  {author} {\bibinfo {author} {\bibfnamefont {N.}~\bibnamefont
		{Majlis}}, \bibinfo {author} {\bibfnamefont {S.}~\bibnamefont {Selzer}}, \
	and\ \bibinfo {author} {\bibfnamefont {G.~C.}\ \bibnamefont {Strinati}},\
}\href {\doibase 10.1103/PhysRevB.45.7872} {\bibfield  {journal} {\bibinfo
	{journal} {Phys. Rev. B}\ }\textbf {\bibinfo {volume} {45}},\ \bibinfo
{pages} {7872} (\bibinfo {year} {1992})}\BibitemShut {NoStop}%
\bibitem [{\citenamefont {Schmidt}\ and\ \citenamefont
	{Thalmeier}(2017)}]{Schmidt214443}%
\BibitemOpen
\bibfield  {author} {\bibinfo {author} {\bibfnamefont {B.}~\bibnamefont
		{Schmidt}}\ and\ \bibinfo {author} {\bibfnamefont {P.}~\bibnamefont
		{Thalmeier}},\ }\href {\doibase 10.1103/PhysRevB.96.214443} {\bibfield
	{journal} {\bibinfo  {journal} {Phys. Rev. B}\ }\textbf {\bibinfo {volume}
		{96}},\ \bibinfo {pages} {214443} (\bibinfo {year} {2017})}\BibitemShut
{NoStop}%
\bibitem [{\citenamefont {Schmidt}\ \emph {et~al.}(2007)\citenamefont
	{Schmidt}, \citenamefont {Thalmeier},\ and\ \citenamefont
	{Shannon}}]{Schmidt125113}%
\BibitemOpen
\bibfield  {author} {\bibinfo {author} {\bibfnamefont {B.}~\bibnamefont
		{Schmidt}}, \bibinfo {author} {\bibfnamefont {P.}~\bibnamefont {Thalmeier}},
	\ and\ \bibinfo {author} {\bibfnamefont {N.}~\bibnamefont {Shannon}},\ }\href
{\doibase 10.1103/PhysRevB.76.125113} {\bibfield  {journal} {\bibinfo
		{journal} {Phys. Rev. B}\ }\textbf {\bibinfo {volume} {76}},\ \bibinfo
	{pages} {125113} (\bibinfo {year} {2007})}\BibitemShut {NoStop}%
\bibitem [{pol()}]{polynomial_fit}%
\BibitemOpen
\href@noop {} {}\bibinfo {note} {The fitting results in
	$a\simeq18.35\times10^{-4}$~mol$^{-1}$K$^{-4}$,
	$b\simeq1.61\times10^{-6}$~mol$^{-1}$K$^{-6}$,
	$c\simeq6.70\times10^{-10}$~mol$^{-1}$K$^{-8}$, and
	$d\simeq9.90\times10^{-14}$~mol$^{-1}$K$^{-10}$}\BibitemShut {NoStop}%
\bibitem [{\citenamefont {Matsumoto}\ \emph {et~al.}(2000)\citenamefont
	{Matsumoto}, \citenamefont {Miyazaki}, \citenamefont {S.~Albrecht},
	\citenamefont {P.~Landee}, \citenamefont {M.~Turnbull},\ and\ \citenamefont
	{Sorai}}]{Matsumoto9993}%
\BibitemOpen
\bibfield  {author} {\bibinfo {author} {\bibfnamefont {T.}~\bibnamefont
		{Matsumoto}}, \bibinfo {author} {\bibfnamefont {Y.}~\bibnamefont {Miyazaki}},
	\bibinfo {author} {\bibfnamefont {A.}~\bibnamefont {S.~Albrecht}}, \bibinfo
	{author} {\bibfnamefont {C.}~\bibnamefont {P.~Landee}}, \bibinfo {author}
	{\bibfnamefont {M.}~\bibnamefont {M.~Turnbull}}, \ and\ \bibinfo {author}
	{\bibfnamefont {M.}~\bibnamefont {Sorai}},\ }\href {\doibase
	10.1021/jp0020081} {\bibfield  {journal} {\bibinfo  {journal} {J. Phys. Chem.
			B}\ }\textbf {\bibinfo {volume} {104}},\ \bibinfo {pages} {9993} (\bibinfo
	{year} {2000})}\BibitemShut {NoStop}%
\bibitem [{\citenamefont {Kobayashi}\ and\ \citenamefont
	{Haseda}(1963)}]{Hanako541}%
\BibitemOpen
\bibfield  {author} {\bibinfo {author} {\bibfnamefont {H.}~\bibnamefont
		{Kobayashi}}\ and\ \bibinfo {author} {\bibfnamefont {T.}~\bibnamefont
		{Haseda}},\ }\href@noop {} {\bibfield  {journal} {\bibinfo  {journal}
		{Journal of the Physical Society of Japan}\ }\textbf {\bibinfo {volume}
		{18}},\ \bibinfo {pages} {541} (\bibinfo {year} {1963})}\BibitemShut
{NoStop}%
\bibitem [{\citenamefont {Yamagata}\ and\ \citenamefont
	{Abe}(1983)}]{Yamagata1179}%
\BibitemOpen
\bibfield  {author} {\bibinfo {author} {\bibfnamefont {K.}~\bibnamefont
		{Yamagata}}\ and\ \bibinfo {author} {\bibfnamefont {H.}~\bibnamefont {Abe}},\
}\href {\doibase http://dx.doi.org/10.1016/0304-8853(83)90852-1} {\bibfield
{journal} {\bibinfo  {journal} {J. Magn. Magn. Mater.}\ }\textbf {\bibinfo
	{volume} {31}},\ \bibinfo {pages} {1179 } (\bibinfo {year}
{1983})}\BibitemShut {NoStop}%
\bibitem [{\citenamefont {Koyama}\ \emph {et~al.}(1987)\citenamefont {Koyama},
	\citenamefont {Nobumasa},\ and\ \citenamefont {Matsuura}}]{Koyama1553}%
\BibitemOpen
\bibfield  {author} {\bibinfo {author} {\bibfnamefont {K.}~\bibnamefont
		{Koyama}}, \bibinfo {author} {\bibfnamefont {H.}~\bibnamefont {Nobumasa}}, \
	and\ \bibinfo {author} {\bibfnamefont {M.}~\bibnamefont {Matsuura}},\ }\href
{\doibase 10.1143/JPSJ.56.1553} {\bibfield  {journal} {\bibinfo  {journal}
		{J. Phys. Soc. Jpn.}\ }\textbf {\bibinfo {volume} {56}},\ \bibinfo {pages}
	{1553} (\bibinfo {year} {1987})}\BibitemShut {NoStop}%
\bibitem [{\citenamefont {Abrahams}(1962)}]{Abrahams56}%
\BibitemOpen
\bibfield  {author} {\bibinfo {author} {\bibfnamefont {S.~C.}\ \bibnamefont
		{Abrahams}},\ }\href {\doibase 10.1063/1.1732318} {\bibfield  {journal}
	{\bibinfo  {journal} {The Journal of Chemical Physics}\ }\textbf {\bibinfo
		{volume} {36}},\ \bibinfo {pages} {56} (\bibinfo {year} {1962})},\ \Eprint
{http://arxiv.org/abs/https://doi.org/10.1063/1.1732318}
{https://doi.org/10.1063/1.1732318} \BibitemShut {NoStop}%
\bibitem [{\citenamefont {Algra}\ \emph {et~al.}(1978)\citenamefont {Algra},
	\citenamefont {de~Jongh},\ and\ \citenamefont {Carlin}}]{Algra24}%
\BibitemOpen
\bibfield  {author} {\bibinfo {author} {\bibfnamefont {H.}~\bibnamefont
		{Algra}}, \bibinfo {author} {\bibfnamefont {L.}~\bibnamefont {de~Jongh}}, \
	and\ \bibinfo {author} {\bibfnamefont {R.}~\bibnamefont {Carlin}},\ }\href
{\doibase http://dx.doi.org/10.1016/0378-4363(78)90107-9} {\bibfield
	{journal} {\bibinfo  {journal} {Physica B+C}\ }\textbf {\bibinfo {volume}
		{93}},\ \bibinfo {pages} {24 } (\bibinfo {year} {1978})}\BibitemShut
{NoStop}%
\bibitem [{\citenamefont {Woodward}\ \emph {et~al.}(2007)\citenamefont
	{Woodward}, \citenamefont {Gibson}, \citenamefont {Jameson}, \citenamefont
	{Landee}, \citenamefont {Turnbull},\ and\ \citenamefont
	{Willett}}]{Woodward4256}%
\BibitemOpen
\bibfield  {author} {\bibinfo {author} {\bibfnamefont {F.~M.}\ \bibnamefont
		{Woodward}}, \bibinfo {author} {\bibfnamefont {P.~J.}\ \bibnamefont
		{Gibson}}, \bibinfo {author} {\bibfnamefont {G.~B.}\ \bibnamefont {Jameson}},
	\bibinfo {author} {\bibfnamefont {C.~P.}\ \bibnamefont {Landee}}, \bibinfo
	{author} {\bibfnamefont {M.~M.}\ \bibnamefont {Turnbull}}, \ and\ \bibinfo
	{author} {\bibfnamefont {R.~D.}\ \bibnamefont {Willett}},\ }\href {\doibase
	10.1021/ic0621392} {\bibfield  {journal} {\bibinfo  {journal} {Inorganic
			Chemistry}\ }\textbf {\bibinfo {volume} {46}},\ \bibinfo {pages} {4256}
	(\bibinfo {year} {2007})}\BibitemShut {NoStop}%
\bibitem [{\citenamefont {Manson}\ \emph {et~al.}(2006)\citenamefont {Manson},
	\citenamefont {Conner}, \citenamefont {Schlueter}, \citenamefont {Lancaster},
	\citenamefont {Blundell}, \citenamefont {Brooks}, \citenamefont {Pratt},
	\citenamefont {Papageorgiou}, \citenamefont {Bianchi}, \citenamefont
	{Wosnitza},\ and\ \citenamefont {Whangbo}}]{Manson4894}%
\BibitemOpen
\bibfield  {author} {\bibinfo {author} {\bibfnamefont {J.~L.}\ \bibnamefont
		{Manson}}, \bibinfo {author} {\bibfnamefont {M.~M.}\ \bibnamefont {Conner}},
	\bibinfo {author} {\bibfnamefont {J.~A.}\ \bibnamefont {Schlueter}}, \bibinfo
	{author} {\bibfnamefont {T.}~\bibnamefont {Lancaster}}, \bibinfo {author}
	{\bibfnamefont {S.~J.}\ \bibnamefont {Blundell}}, \bibinfo {author}
	{\bibfnamefont {M.~L.}\ \bibnamefont {Brooks}}, \bibinfo {author}
	{\bibfnamefont {F.~L.}\ \bibnamefont {Pratt}}, \bibinfo {author}
	{\bibfnamefont {T.}~\bibnamefont {Papageorgiou}}, \bibinfo {author}
	{\bibfnamefont {A.~D.}\ \bibnamefont {Bianchi}}, \bibinfo {author}
	{\bibfnamefont {J.}~\bibnamefont {Wosnitza}}, \ and\ \bibinfo {author}
	{\bibfnamefont {M.-H.}\ \bibnamefont {Whangbo}},\ }\href {\doibase
	10.1039/B608791D} {\bibfield  {journal} {\bibinfo  {journal} {Chem. Commun.}\
	}\textbf {\bibinfo {volume} {0}},\ \bibinfo {pages} {4894} (\bibinfo {year}
	{2006})}\BibitemShut {NoStop}%
\bibitem [{\citenamefont {Woodward}\ \emph {et~al.}(2002)\citenamefont
	{Woodward}, \citenamefont {Albrecht}, \citenamefont {Wynn}, \citenamefont
	{Landee},\ and\ \citenamefont {Turnbull}}]{Woodward144412}%
\BibitemOpen
\bibfield  {author} {\bibinfo {author} {\bibfnamefont {F.~M.}\ \bibnamefont
		{Woodward}}, \bibinfo {author} {\bibfnamefont {A.~S.}\ \bibnamefont
		{Albrecht}}, \bibinfo {author} {\bibfnamefont {C.~M.}\ \bibnamefont {Wynn}},
	\bibinfo {author} {\bibfnamefont {C.~P.}\ \bibnamefont {Landee}}, \ and\
	\bibinfo {author} {\bibfnamefont {M.~M.}\ \bibnamefont {Turnbull}},\ }\href
{\doibase 10.1103/PhysRevB.65.144412} {\bibfield  {journal} {\bibinfo
		{journal} {Phys. Rev. B}\ }\textbf {\bibinfo {volume} {65}},\ \bibinfo
	{pages} {144412} (\bibinfo {year} {2002})}\BibitemShut {NoStop}%
\bibitem [{\citenamefont {Nath}\ \emph
	{et~al.}(2008{\natexlab{b}})\citenamefont {Nath}, \citenamefont {Tsirlin},
	\citenamefont {Kaul}, \citenamefont {Baenitz}, \citenamefont {B\"uttgen},
	\citenamefont {Geibel},\ and\ \citenamefont {Rosner}}]{Nath024418}%
\BibitemOpen
\bibfield  {author} {\bibinfo {author} {\bibfnamefont {R.}~\bibnamefont
		{Nath}}, \bibinfo {author} {\bibfnamefont {A.~A.}\ \bibnamefont {Tsirlin}},
	\bibinfo {author} {\bibfnamefont {E.~E.}\ \bibnamefont {Kaul}}, \bibinfo
	{author} {\bibfnamefont {M.}~\bibnamefont {Baenitz}}, \bibinfo {author}
	{\bibfnamefont {N.}~\bibnamefont {B\"uttgen}}, \bibinfo {author}
	{\bibfnamefont {C.}~\bibnamefont {Geibel}}, \ and\ \bibinfo {author}
	{\bibfnamefont {H.}~\bibnamefont {Rosner}},\ }\href {\doibase
	10.1103/PhysRevB.78.024418} {\bibfield  {journal} {\bibinfo  {journal} {Phys.
			Rev. B}\ }\textbf {\bibinfo {volume} {78}},\ \bibinfo {pages} {024418}
	(\bibinfo {year} {2008}{\natexlab{b}})}\BibitemShut {NoStop}%
\bibitem [{\citenamefont {Bernu}\ and\ \citenamefont
	{Misguich}(2001)}]{Bernu134409}%
\BibitemOpen
\bibfield  {author} {\bibinfo {author} {\bibfnamefont {B.}~\bibnamefont
		{Bernu}}\ and\ \bibinfo {author} {\bibfnamefont {G.}~\bibnamefont
		{Misguich}},\ }\href {\doibase 10.1103/PhysRevB.63.134409} {\bibfield
	{journal} {\bibinfo  {journal} {Phys. Rev. B}\ }\textbf {\bibinfo {volume}
		{63}},\ \bibinfo {pages} {134409} (\bibinfo {year} {2001})}\BibitemShut
{NoStop}%
\bibitem [{\citenamefont {Hofmann}\ \emph {et~al.}(2003)\citenamefont
	{Hofmann}, \citenamefont {Lorenz}, \citenamefont {Berggold}, \citenamefont
	{Gr\"uninger}, \citenamefont {Freimuth}, \citenamefont {Uhrig},\ and\
	\citenamefont {Br\"uck}}]{Hofmann184502}%
\BibitemOpen
\bibfield  {author} {\bibinfo {author} {\bibfnamefont {M.}~\bibnamefont
		{Hofmann}}, \bibinfo {author} {\bibfnamefont {T.}~\bibnamefont {Lorenz}},
	\bibinfo {author} {\bibfnamefont {K.}~\bibnamefont {Berggold}}, \bibinfo
	{author} {\bibfnamefont {M.}~\bibnamefont {Gr\"uninger}}, \bibinfo {author}
	{\bibfnamefont {A.}~\bibnamefont {Freimuth}}, \bibinfo {author}
	{\bibfnamefont {G.~S.}\ \bibnamefont {Uhrig}}, \ and\ \bibinfo {author}
	{\bibfnamefont {E.}~\bibnamefont {Br\"uck}},\ }\href {\doibase
	10.1103/PhysRevB.67.184502} {\bibfield  {journal} {\bibinfo  {journal} {Phys.
			Rev. B}\ }\textbf {\bibinfo {volume} {67}},\ \bibinfo {pages} {184502}
	(\bibinfo {year} {2003})}\BibitemShut {NoStop}%
\bibitem [{\citenamefont {Oitmaa}\ and\ \citenamefont
	{Bornilla}(1996)}]{Oitmaa14228}%
\BibitemOpen
\bibfield  {author} {\bibinfo {author} {\bibfnamefont {J.}~\bibnamefont
		{Oitmaa}}\ and\ \bibinfo {author} {\bibfnamefont {E.}~\bibnamefont
		{Bornilla}},\ }\href {\doibase 10.1103/PhysRevB.53.14228} {\bibfield
	{journal} {\bibinfo  {journal} {Phys. Rev. B}\ }\textbf {\bibinfo {volume}
		{53}},\ \bibinfo {pages} {14228} (\bibinfo {year} {1996})}\BibitemShut
{NoStop}%
\bibitem [{\citenamefont {Johnston}\ \emph {et~al.}(2000)\citenamefont
	{Johnston}, \citenamefont {Kremer}, \citenamefont {Troyer}, \citenamefont
	{Wang}, \citenamefont {Kl\"umper}, \citenamefont {Bud'ko}, \citenamefont
	{Panchula},\ and\ \citenamefont {Canfield}}]{Johnston9558}%
\BibitemOpen
\bibfield  {author} {\bibinfo {author} {\bibfnamefont {D.~C.}\ \bibnamefont
		{Johnston}}, \bibinfo {author} {\bibfnamefont {R.~K.}\ \bibnamefont
		{Kremer}}, \bibinfo {author} {\bibfnamefont {M.}~\bibnamefont {Troyer}},
	\bibinfo {author} {\bibfnamefont {X.}~\bibnamefont {Wang}}, \bibinfo {author}
	{\bibfnamefont {A.}~\bibnamefont {Kl\"umper}}, \bibinfo {author}
	{\bibfnamefont {S.~L.}\ \bibnamefont {Bud'ko}}, \bibinfo {author}
	{\bibfnamefont {A.~F.}\ \bibnamefont {Panchula}}, \ and\ \bibinfo {author}
	{\bibfnamefont {P.~C.}\ \bibnamefont {Canfield}},\ }\href {\doibase
	10.1103/PhysRevB.61.9558} {\bibfield  {journal} {\bibinfo  {journal} {Phys.
			Rev. B}\ }\textbf {\bibinfo {volume} {61}},\ \bibinfo {pages} {9558}
	(\bibinfo {year} {2000})}\BibitemShut {NoStop}%
\bibitem [{\citenamefont {de~Jongh}(2012)}]{Jongh2012}%
\BibitemOpen
\bibfield  {author} {\bibinfo {author} {\bibfnamefont {L.~J.}\ \bibnamefont
		{de~Jongh}},\ }\href@noop {} {\emph {\bibinfo {title} {Magnetic properties of
			layered transition metal compounds}}},\ Vol.~\bibinfo {volume} {9}\ (\bibinfo
{publisher} {Springer},\ \bibinfo {address} {Dordrecht},\ \bibinfo {year}
{2012})\BibitemShut {NoStop}%
\bibitem [{\citenamefont {Nath}\ \emph {et~al.}(2014)\citenamefont {Nath},
	\citenamefont {Ranjith}, \citenamefont {Roy}, \citenamefont {Johnston},
	\citenamefont {Furukawa},\ and\ \citenamefont {Tsirlin}}]{Nath024431}%
\BibitemOpen
\bibfield  {author} {\bibinfo {author} {\bibfnamefont {R.}~\bibnamefont
		{Nath}}, \bibinfo {author} {\bibfnamefont {K.~M.}\ \bibnamefont {Ranjith}},
	\bibinfo {author} {\bibfnamefont {B.}~\bibnamefont {Roy}}, \bibinfo {author}
	{\bibfnamefont {D.~C.}\ \bibnamefont {Johnston}}, \bibinfo {author}
	{\bibfnamefont {Y.}~\bibnamefont {Furukawa}}, \ and\ \bibinfo {author}
	{\bibfnamefont {A.~A.}\ \bibnamefont {Tsirlin}},\ }\href {\doibase
	10.1103/PhysRevB.90.024431} {\bibfield  {journal} {\bibinfo  {journal} {Phys.
			Rev. B}\ }\textbf {\bibinfo {volume} {90}},\ \bibinfo {pages} {024431}
	(\bibinfo {year} {2014})}\BibitemShut {NoStop}%
\bibitem [{\citenamefont {Eisele}\ and\ \citenamefont
	{Keffer}(1954)}]{Eisele929}%
\BibitemOpen
\bibfield  {author} {\bibinfo {author} {\bibfnamefont {J.~A.}\ \bibnamefont
		{Eisele}}\ and\ \bibinfo {author} {\bibfnamefont {F.}~\bibnamefont
		{Keffer}},\ }\href {\doibase 10.1103/PhysRev.96.929} {\bibfield  {journal}
	{\bibinfo  {journal} {Phys. Rev.}\ }\textbf {\bibinfo {volume} {96}},\
	\bibinfo {pages} {929} (\bibinfo {year} {1954})}\BibitemShut {NoStop}%
\bibitem [{\citenamefont {Kitaoka}\ \emph {et~al.}(1998)\citenamefont
	{Kitaoka}, \citenamefont {Kobayashi}, \citenamefont {Kōda}, \citenamefont
	{Wakabayashi}, \citenamefont {Niino}, \citenamefont {Yamakage}, \citenamefont
	{Taguchi}, \citenamefont {Amaya}, \citenamefont {Yamaura}, \citenamefont
	{Takano}, \citenamefont {Hirano},\ and\ \citenamefont {Kanno}}]{Kitaoka3703}%
\BibitemOpen
\bibfield  {author} {\bibinfo {author} {\bibfnamefont {Y.}~\bibnamefont
		{Kitaoka}}, \bibinfo {author} {\bibfnamefont {T.}~\bibnamefont {Kobayashi}},
	\bibinfo {author} {\bibfnamefont {A.}~\bibnamefont {Kōda}}, \bibinfo
	{author} {\bibfnamefont {H.}~\bibnamefont {Wakabayashi}}, \bibinfo {author}
	{\bibfnamefont {Y.}~\bibnamefont {Niino}}, \bibinfo {author} {\bibfnamefont
		{H.}~\bibnamefont {Yamakage}}, \bibinfo {author} {\bibfnamefont
		{S.}~\bibnamefont {Taguchi}}, \bibinfo {author} {\bibfnamefont
		{K.}~\bibnamefont {Amaya}}, \bibinfo {author} {\bibfnamefont
		{K.}~\bibnamefont {Yamaura}}, \bibinfo {author} {\bibfnamefont
		{M.}~\bibnamefont {Takano}}, \bibinfo {author} {\bibfnamefont
		{A.}~\bibnamefont {Hirano}}, \ and\ \bibinfo {author} {\bibfnamefont
		{R.}~\bibnamefont {Kanno}},\ }\href {\doibase 10.1143/JPSJ.67.3703}
{\bibfield  {journal} {\bibinfo  {journal} {J. Phys. Soc. Jpn.}\ }\textbf
	{\bibinfo {volume} {67}},\ \bibinfo {pages} {3703} (\bibinfo {year}
	{1998})}\BibitemShut {NoStop}%
\bibitem [{\citenamefont {Ramirez}\ \emph {et~al.}(1990)\citenamefont
	{Ramirez}, \citenamefont {Espinosa},\ and\ \citenamefont
	{Cooper}}]{Ramirez2070}%
\BibitemOpen
\bibfield  {author} {\bibinfo {author} {\bibfnamefont {A.~P.}\ \bibnamefont
		{Ramirez}}, \bibinfo {author} {\bibfnamefont {G.~P.}\ \bibnamefont
		{Espinosa}}, \ and\ \bibinfo {author} {\bibfnamefont {A.~S.}\ \bibnamefont
		{Cooper}},\ }\href {\doibase 10.1103/PhysRevLett.64.2070} {\bibfield
	{journal} {\bibinfo  {journal} {Phys. Rev. Lett.}\ }\textbf {\bibinfo
		{volume} {64}},\ \bibinfo {pages} {2070} (\bibinfo {year}
	{1990})}\BibitemShut {NoStop}%
\end{thebibliography}

%

\end{document}